\documentclass[a4paper,12pt]{amsart}
\newtheorem{theoreme}{Theorem}[section]

\newtheorem{lemme}{Lemma}[section]
\newtheorem{corollaire}{Corollary}[section]
\theoremstyle{remark}
\newtheorem{remarque}{Remark}[section]
\newtheorem{remarques}[remarque]{Remarks}
\numberwithin{equation}{section}
\renewcommand{\tilde}{\widetilde}
\newcommand{\iref}[1]{(\ref{#1})}
\newcommand{\nl}[2]{\|{#1}\|_{L^{{#2}}}}
\newcommand{\nld}[3]{\|{#1}\|_{L^{{#2}}({#3})}}
\newcommand{\ep}{\varepsilon}
\newcommand{\wt}{\widetilde}
\renewcommand{\t}{\tau}

\newcommand{\Z}{\mathbb{Z}}
\begin{document}
\title{Some results on the Navier-Stokes equations in thin 3D~domains}
\author{Drago\c s Iftimie \and Genevi\`eve Raugel}

\begin{abstract}
We consider the Navier-Stokes equations on thin 3D domains $Q_\ep=\Omega \times
(0,\varepsilon)$, supplemented  mainly with purely periodic boundary conditions or 
with periodic boundary conditions in the
thin direction and homogeneous Dirichlet conditions on the lateral
boundary. We prove global existence and uniqueness of solutions for
initial data and forcing terms, which are
larger and less regular than in previous works. An important
tool in the proofs are some
Sobolev embeddings into anisotropic $L^p$-type
spaces. As in \cite{RaugelSell94},
better results are proved in the purely periodic case, where 
the conservation of enstrophy property is used. For
example, in the case of a vanishing forcing term,
we prove global existence and uniqueness of solutions if
$\|(I-M)u_0\|_{H^{1/2}(Q_\ep)}\exp(C^{-1}
\ep^{-1/s}\|Mu_0\|^{2/s}_{L^2(Q_\ep)})\leq C$
for both boundary conditions or
$\|Mu_0\|_{H^1(Q_\ep)}\leq C\ep^{-\beta}$, $\|(Mu_0)_3\|_{L^2(Q_\ep)}\leq
C\ep^\beta$, $\|(I-M)u_0\|_{H^{1/2}(Q_\ep)}\leq C\ep^{1/4-\beta/2}$ for
purely periodic boundary conditions, where $1/2\leq s<1$ and
$0\leq\beta\leq 1/2$ are arbitrary, $C$ is a prescribed positive 
constant independent of $\ep$ and
$M$ denotes the average operator in the thin direction. We also give a 
new uniqueness criterium for weak Leray solutions. 
\end{abstract}

\maketitle

\begin{flushleft}
\textit{Dedicated to Jack Hale on the occasion of his 70th birthday}
\end{flushleft}
\vskip 10mm

\section{Introduction}

As is well-known, the Navier-Stokes equations describe
the time evolution of solutions of mathematical models of viscous
incompressible fluids. From the mathematical point of view, global
existence of weak solutions is known to hold in every space 
dimension. Uniqueness of weak solutions is known in dimension 2 
(see \cite{Leray}). In dimension 3, to obtain global existence and
uniqueness, one has to assume additional
regularity and smallness assumptions on the initial data and the 
forcing term. A natural
question is how to use the good properties of the 2D Navier-Stokes equations to
improve the uniqueness and regularity results for the 3D equations, when the domain 
is thin. In this paper, we consider the existence and
uniqueness of solutions of the Navier-Stokes equations in
thin three-dimensional domains $Q_\varepsilon = \Omega \times
(0,\varepsilon)$, where $\Omega$ is a suitable bounded domain in
${\mathbb R }^2$ and $\varepsilon$ is a small positive parameter,
$0< \varepsilon <1$. We do a detailed study of this question in
the case of two types of boundary conditions: the
purely periodic condition (PP) and
the periodic-Dirichlet boundary condition (PD), that is, periodic condition
in the thin vertical direction and homogeneous Dirichlet conditions on the
lateral boundary $\Gamma_l = \Omega \times (0,\varepsilon)$. When
(PD) boundary conditions are considered, we assume that $\Omega$ is a
regular domain in ${\mathbb R }^2$, while, in the case of the (PP) boundary
conditions, $\Omega = (0,l_1) \times (0,l_2)$, where $l_1$, $l_2$ are
positive numbers. Our results also hold for other types of boundary conditions, such as
those considered in \cite{TemamZiane96} (See remarks \ref{Remark4}, 
\ref{Remark5} and \ref{RemarkFP}). 
\vskip 2mm
The study of the Navier-Stokes equations on thin domains originates
in a series of papers of Hale and Raugel (\cite{HaleRaugel929292},
\cite{HaleRaugel92929292}, \cite{HaleRaugel9292}), concerning the
reaction-diffusion and damped wave equations on thin domains. 
In thin three-dimensional domains, inspired by
the methods developed in (\cite{HaleRaugel929292},
\cite{HaleRaugel92929292}, \cite{HaleRaugel9292}),
Raugel and Sell (\cite{RaugelSell93}, \cite{RaugelSell94}) 
proved global existence of strong solutions for large
initial data and forcing terms, in the case of the boundary 
conditions (PP) and (PD). As in \cite{HaleRaugel929292},
an essential tool in their proof is
the vertical mean operator M (see (\ref{meanthin})), which allows
to decompose every function $g$ on $Q_{\varepsilon}$ into the sum
of a function $Mg$
which does not depend on the vertical variable and a function
$(I-M)g$ with vanishing vertical mean and thus to use more precise
Sobolev and Poincar\'{e} inequalities. Later, in the case of Dirichlet
boundary conditions, Avrin \cite{Avrin} showed global
existence of strong solutions of the Navier-Stokes equations on
thin three-dimensional domains for large data, by applying a contraction
principle argument and carefully analyzing the dependence of the
solution on the first eigenvalue of the corresponding
Laplace operator. The analysis in the case of Dirichlet boundary conditions in a thin
domain is simpler, because the size of the first eigenvalue is of
order $\varepsilon^{-2}$ and thus the above decomposition is of no use.
Next, using the same tools as Raugel and Sell
as well as  improved Agmon inequalities, Temam and Ziane (\cite{TemamZiane96},
\cite{TemamZiane97}) generalized the results of \cite{RaugelSell93}, 
\cite{RaugelSell94} to other boundary conditions
and, in the case of the free boundary conditions, to thin spherical
domains. In the periodic case, Moise, Temam and Ziane (see 
\cite{MoiseTemamZiane}) proved global existence of strong solutions 
for initial data, that are larger than in \cite{RaugelSell94}. Also in the
(PP) case, using
anisotropic spaces, Iftimie \cite{Iftimie9999} showed
existence and uniqueness of solutions for less regular initial data
and proved that initial data $u_0$ with larger $(I-M)u_0$ part
could be taken.  Finally, in the same case, Montgomery-Smith
\cite{MontgomerySmith}
gives global existence results, which are not contained in
\cite{MoiseTemamZiane}.

In this paper, we improve the previous existence and uniqueness
results in two directions, by requiring less regularity on the initial
data and by allowing a larger size of the initial data and forcing term.
We also emphasize the importance played by the third component of the
vertical mean value of the data. For instance, in the (PD) and (PP)
cases, we show that, for any real number $\gamma$, $0 \leq 
\gamma<1/2$, there exists a positive constant $K_\gamma$ such that, 
for $0 < \varepsilon \leq 1$, if the initial data $u_0$ and forcing term $f$ satisfy
\begin{equation*}
 \begin{split}
 &\|Mu_0\|_{L^2(Q_\varepsilon)} \, \leq \, K_{\gamma} \varepsilon^{1/2}, 
 \quad\|A^{1/4}_{\varepsilon}(I-M)u_0\|_{L^2(Q_\varepsilon)} \, \leq \,
K_{\gamma},\\
 &\sup_t \|MP_\varepsilon f(t)\|_{L^2(Q_\varepsilon)} \, \leq \, K_{\gamma}
 \varepsilon^{1/2}, 
\quad\sup_t \|(I-M)P_\varepsilon f(t)\|_{L^2
 (Q_\varepsilon)} \, \leq \, K_{\gamma}
 \varepsilon^{-1/2}|\ln\varepsilon|^{\gamma},
  \end{split}
\end{equation*}
where $A_{\varepsilon}$ is the Stokes operator and $P_{\varepsilon}$
is the Leray projection, then the Navier-Stokes equations have a
global solution $u \in C^0([0, + \infty); (L^2(Q_\varepsilon))^3)$, which is
unique in the class of weak Leray solutions. In the purely periodic 
case, one can also choose $\gamma=1/2$; furthermore, in
this case, assuming that $Mu_0$ is more regular, we 
obtain global existence of
a solution $u$ in $C^0([0, + \infty); (H^{1/2}(Q_\varepsilon))^3)$,
which is unique in the class of weak Leray solutions, if, 
for instance, $u_0$ and $f$ satisfy
{\allowdisplaybreaks \begin{gather*}
 \|A^{1/2}_\varepsilon(Mu_0)\|_{L^2(Q_\varepsilon)} +
\sup_t \|MP_\varepsilon f(t)\|_{L^2(Q_\varepsilon)} \, \leq
\, k_0 \varepsilon^{-\beta}~,\\
 \|Mu_{03}\|_{L^2(Q_\varepsilon)}
+ \sup_t \|A^{-1/2 }_\varepsilon(MP_\varepsilon f_3(t))\|_{L^2(Q_\varepsilon)}
\, \leq \, k_0 \varepsilon^\beta~,\\
\sup_t \|(I-M)P_\varepsilon f(t)\|_{L^2(Q_\varepsilon)}\, \leq \, k_0
\varepsilon^{-3/4 -\beta/2} \, ,\\
  \|A^{1/4}_\varepsilon((I-M)u_0)\|_{L^2(Q_\varepsilon)}
 \, \leq \, k_0 \varepsilon^{1/4 -\beta/2}~,
 \end{gather*}}
where $0 \leq \beta \leq 1/2$.
These results are stated more precisely in the theorems
\ref{theoreme1}, \ref{theoreme1bis} and \ref{theoreme2} below. To
obtain these existence results, we at first show sharp estimates of
the nonlinear term appearing in the Navier-Stokes equations by working
in the anisotropic Sobolev spaces $L^{p,p'}(Q_\varepsilon)=
L^p(\Omega;L^{p'}(0,\varepsilon))$, for $p \neq p'$ and also by
taking into account commutator properties. In the purely periodic case,
like in \cite{RaugelSell94}, we use the conservation of
enstrophy
of the variable $M \tilde{u}(t)= (Mu_1(t),Mu_2(t),0)$. But, unlike
\cite{RaugelSell94}, we
work directly in the domain $Q_\varepsilon$, that is, we do not rescale
the domain $Q_\varepsilon$ to a domain of thickness $1$. 

\vskip 2mm

We recall that the Navier-Stokes equations in the bounded domain
$Q_\varepsilon$ are given by
\begin{equation}
\label{NS}
\begin{split}
&\partial_t u -\nu \Delta u + (u \cdot \nabla)u +\nabla p\, =\, f~, \\
& {\rm div}u \, =\, 0~,\\
& u( \cdot, 0)\, =\, u_0~,
\end{split}
\end{equation}
where $\nabla$ is the gradient operator, $\Delta$ is the Laplace
operator, $f$ is a forcing term and $u(x,t)=(u_1,u_2,u_3)(x,t)$,
$p(x,t)$ are the velocity vector and the pressure at point $x=(x_1,x_2,x_3)$
and time $t$ respectively. We assume that the viscosity $\nu$ is a
fixed positive number.
Here these equations are mainly supplemented either with
the periodic-Dirichlet boundary conditions (PD) or with purely periodic 
conditions (PP) on $\partial Q_{\varepsilon}$.  In the (PP)
case, we require in addition that the data $u_0$ and $f$ have a
vanishing total mean value, that is,
\begin{equation}
\label{moyennetot}
 \int_{Q_{\varepsilon}} u_0 dx\, = \, \int_{Q_{\varepsilon}} f
 dx\,=\,0~.
\end{equation}

In order to describe our results more precisely and write the
Navier-Stokes equations in an abstract form, we need to introduce
some notation.  For $m \in {\mathbb N}$, we denote by
$H^m(Q_\varepsilon)$ the Hilbert space $\{g \in L^2(Q_{\varepsilon});
\sum_{0\leq |j| \leq m} (\int_{Q_{\varepsilon}} |D^j g|^2 dx) < +
\infty \}$ equipped with the classical norm $\|\cdot\|_{H^m}$. For
$m<s<m+1$, we denote by $H^s(Q_{\varepsilon})$ the interpolated
Hilbert space $[H^m(Q_{\varepsilon}),
H^{m+1}(Q_{\varepsilon})]_{\theta}$, where $\theta =s-m$ and we endow
this space with the standard norm $\|\cdot\|_{H^s}$.
As usual, $H^0(Q_\varepsilon)$
is denoted by $L^2(Q_\varepsilon)$ and
$\|g\|_{L^2} = (\int_{Q_{\varepsilon}} g^2 dx)^{1/2}$.
Likewise, for $m\geq 0$, we introduce the
space $H^m_p(Q_\varepsilon)$, which is the closure in $H^m(Q_\varepsilon)$
of those smooth functions that are periodic in $Q_\varepsilon$,
and, for $m<s<m+1$, we introduce the interpolated Hilbert space
$H^s_p(Q_\varepsilon)= [H^m_p(Q_{\varepsilon}),
H^{m+1}_p(Q_{\varepsilon})]_{\theta}$, where $\theta =s-m$.
We also define the spaces
$\dot{H}^s_p(Q_\varepsilon) =\left\{g \in H^s_p(Q_\varepsilon);
\int_{Q_{\varepsilon}} g(x) dx \, = \, 0 \right\}$.
The spaces $H^s_p(Q_\varepsilon)$ and $\dot{H}^s_p(Q_\varepsilon)$
can be described in terms of Fourier series; for $k$ in the integer
lattice ${\mathbb Z }^3$, we set $ka \equiv (k_1a_1,
k_2a_2,k_3a_3)$, where $a_1=l^{-1}_1$, $a_2=l^{-1}_2$,
$a_3=\varepsilon^{-1}$,
and we write
\begin{equation}
\label{Fourier}
g(x)\, =\, \varepsilon^{-1/2}\sqrt{a_1a_2}
\sum_{k \in {\mathbb Z }^3} g_k \exp(2i\pi ka \cdot x)~,
\end{equation}
where $g_k \in {\mathbb R }$, $\overline{g}_k=g_{-k}$ and $g_k=
\varepsilon^{-1/2}\sqrt{a_1a_2} \int_{Q_{\varepsilon}} g(x)
\exp(-2i\pi ka\cdot x) dx$.  Then,
$g \in H^s_p(Q_\varepsilon)$ is in the subspace
$\dot{H}^s_p(Q_\varepsilon)$ if and only if $g_{(0,0,0)}=0$.  The
usual norm $\|g\|_{H^s}$ and semi-norm $|g|_{H^s}$ on
$H^s_p(Q_\varepsilon)$ can be expressed as follows
\begin{equation}
\|g\|^2_{H^s}\,=\, \sum_{k \in {\mathbb Z }^3} (1+
|ka|^2)^s|g_k|^2\, , \quad  |g|^2_{H^s}\,=\, \sum_{k
\in {\mathbb Z }^3} |ka|^{2s}|g_k|^2~,
\end{equation}
and the semi-norm $|\cdot|_{H^s}$ is actually a norm on the subspace
$\dot{H}^s_p(Q_\varepsilon)$. We now define the operator $\Delta_p
=\Delta$, with domain $D(-\Delta_p) = \dot{H}^2_p(Q_\varepsilon) $.
Clearly,
for $0 \leq s\leq 2$, $D((-\Delta_p)^{s/2}) =\dot{H}^{s}_p(Q_\varepsilon)$
and the semi-norm $\|\cdot\|_s \equiv \|(-\Delta_p)^{s/2})
\cdot\|_{L^2}$ is a norm on $\dot{H}^{s}_p(Q_\varepsilon)$, which is
equivalent to the norm $|\cdot|_{H^s}$ , with constants independent 
of $\varepsilon$.

In the (PD) boundary case, we introduce the space $H^1_d(Q_\varepsilon)$,
which is the closure in $H^1(Q_\varepsilon)$ of those smooth functions that
are periodic, of period $\varepsilon$ in the vertical direction and
have compact support in $\Omega \times [0,\varepsilon]$.
We then define the operator $\Delta_d
=\Delta$, with domain $D(-\Delta_d) = \{g \in H^1_d(Q_\varepsilon);
\Delta g \in L^2(Q_\varepsilon) \}$. Clearly, $D((-\Delta_d)^{1/2}) =
H^1_d(Q_\varepsilon)$. For $0 \leq s \leq 2$, we thus introduce the
space $H^{s}_d(Q_\varepsilon) \equiv D((-\Delta_d)^{s/2})$ equipped
with the graph norm $\|\cdot\|_s \equiv \|(-\Delta_d)^{s/2}) \cdot\|_{L^2}$.
We recall
that, for $0 \leq s < 1/2$, $H^{s}_d(Q_\varepsilon) =
H^{s}(Q_\varepsilon)$ and that, for $1/2 < s \leq 2$,
$H^{s}_d(Q_\varepsilon) = \{g \in H^s(Q_\varepsilon); g=0~ {\rm on}~
\Gamma_l~; ~g~ {\rm periodic~in ~the ~ variable~ } x_3 \}$ (see
\cite{Grisvard}). The case $s=1/2$ is more delicate but here we do not need
to characterize this space. For details on this question, we refer to
\cite{LionsMagenes}. 

Below, when there is no confusion, we denote by $X^s$ the space
$\dot{H}^{s}_p(Q_\varepsilon)$ or $H^{s}_d(Q_\varepsilon)$. Using
Fourier series in the vertical direction or arguing as in
\cite{MarsdenRatiuRaugel}, one shows that there there exists a
positive constant $c_0 \geq 1$ such that, for all $g \in X^2$
\begin{equation}
\label{equivnorm2}
c^{-1}_0 \|g\|_2 \, \leq \, ( \sum_{j=0}^{j=2} \|D^j g\|^2_{L^2})^{1/2}\,
\leq \, c_0\|g\|_2~,
\end{equation}
which implies, by interpolation, that there exists a constant $c_1$ such that,
for all $g \in X^s$ with $0 \leq s \leq 2$,
\begin{equation}
\label{inegnorms}
 \|g\|_{H^s} \, \leq \, c_1\|g\|_s~,
\end{equation}

Since we are dealing here with vectors, we introduce the
spaces $(H^{s}(Q_\varepsilon))^3$, $(H^{s}_d(Q_\varepsilon))^3$,
$(H^{s}_p(Q_\varepsilon))^3$ as well , etc..., equipped with the corresponding
norms and semi-norms. For the abstract setting of the Navier-Stokes
equations, we classically consider a Hilbert space $H_{\varepsilon}$,
which is a subspace of $(L^2(Q_\varepsilon))^3$ and depends on the
boundary conditions. In the (PP) case, $H_\varepsilon = H_p$ denotes
the closure in $(L^2(Q_\varepsilon))^3$ of those smooth vectors $u$
that are periodic in $Q_\varepsilon$ and satisfy
\begin{equation}
\int_{Q_{\varepsilon}} u(x) dx \, = \, 0\, , \quad {\rm div}u\,=\,0~.
\end{equation}
In the (PD) case, $H_\varepsilon = H_d$ denotes
the closure in $(L^2(Q_\varepsilon))^3$ of those smooth vectors $u$
that are periodic in the vertical direction, have compact support in
$\Omega \times [0,\varepsilon]$ and satisfy ${\rm div}u=0$ in
$Q_\varepsilon$. The classical subspaces
\begin{equation*}
\begin{split}
 &V_\varepsilon \, = \, V_p \, \equiv \, H_p \cap
 (\dot{H}^1_p(Q_\varepsilon) )^3\, =
 \, \left\{u \in (\dot{H}^1_p(Q_\varepsilon))^3;
 {\rm div}u=0 \right\}~, \\
 & V_\varepsilon \, =\,  V_d \, \equiv \, H_d \cap
 ({H}^1_d(Q_\varepsilon))^3 \,  = \, \left\{u \in
 (H^1_d(Q_\varepsilon))^3; {\rm
 div}u=0 \right\}~,
\end{split}
\end{equation*}
are also useful.
If $((\cdot,\cdot))$ denotes the inner product on $V_\varepsilon$, we
introduce the Stokes operator $A_\varepsilon$ as the isomorphism from
$V_\varepsilon$ onto the dual $V'_\varepsilon$ of $V_\varepsilon$ defined by
\begin{equation*}
\langle A_\varepsilon u , v \rangle_{V'_\varepsilon, V_\varepsilon} \, =\,
((u, v))\, , \quad \forall v \in V_\varepsilon~.
\end{equation*}
One can also extend $A_\varepsilon$ as a linear unbounded operator on
$H_\varepsilon$.  The domain $D(A_\varepsilon) \equiv
 \left\{u \in
V_\varepsilon; A_\varepsilon u \in H_\varepsilon\right\}$ is exactly
the space $(H^2(Q_\varepsilon))^3 \cap V_\varepsilon$, in the (PP) and
(PD) cases that we consider here.  If $P_\varepsilon$ denotes the
orthogonal (Leray) projection of $(L^2(Q_\varepsilon))^3$ onto
$H_\varepsilon$, the Stokes operator $A_\varepsilon$ is given by
\begin{equation*}
A_\varepsilon u \,= \, -P_\varepsilon \Delta u \, , \quad \forall u \in
D(A_\varepsilon)~.
\end{equation*}
Furthermore, in the cases (PP) and (PD), the
Cattabriga-Solonnikov inequality holds uniformly in $\varepsilon$,
that is, there exist  positive constants $c_2 = c_2(\Omega) > 1$ and
$c_3 = c_3(\Omega) > 1$, such that, for $0<\varepsilon \leq 1$, for
any $u \in D(A_\varepsilon)$,
\begin{equation}
\label{Catta}
c^{-1}_3 ( \sum_{j=0}^{j=2} \|D^j u\|^2_{L^2})^{1/2} \, \leq \,
c^{-1}_2 \|\Delta u \|_{L^2} \, \leq \,
 \|A_\varepsilon u\|_{L^2}\, \leq \, c_2 \|\Delta u \|_{L^2} \, \leq \,
c_3 ( \sum_{j=0}^{j=2} \|D^j u\|^2_{L^2})^{1/2} ~.
\end{equation}
In the (PP) case, the property (\ref{Catta}) directly follows from
(\ref{equivnorm2}), since then $A_\varepsilon u =  - \Delta u $, for all $u
\in D(A_\varepsilon)$. In the (PD) case, the inequality (\ref{Catta})
is proved, as in \cite{MarsdenRatiuRaugel}, by extending $u$ by
periodicity to the domain $Q_1 = \Omega \times [0,1]$ and applying
the known Cattabriga-Solonnikov inequality in $Q_1$.

For $0 \leq s \leq 2$, we denote by $V^s_\varepsilon$ the space
$D(A^{s/2}_\varepsilon)$, equipped with the natural norm $\|
\cdot\|_{V^s_\varepsilon}
\equiv |\cdot|_s \equiv \|A^{s/2}_\varepsilon \cdot\|
_{L^2}$. Arguing as in \cite{FujitaMorimoto} and using (\ref{Catta}),
one shows that, for $0 \leq s\leq 2$, $D(A^{s/2}_\varepsilon) =
(X^s)^3 \cap H_{\varepsilon}$ and that
there exists a constant $c_4 >1$, such that, for $0 \leq s \leq 2$,
\begin{equation}
\label{Cattas}
c^{-1}_4 \|(-\Delta)^{s/2} u \|_{L^2} \, \leq \,
 \|A^{s/2}_\varepsilon u\|_{L^2}\, \leq \, c_4 \|(-\Delta)^{s/2} u \|_{L^2} \,
 , \quad \forall u \in V^s_\varepsilon~.
\end{equation}
For $0\leq s \leq 1$, we also consider the dual space $V^{-s}_\varepsilon
\equiv D(A^{-s/2}_\varepsilon)$ of $D(A^{s/2}_\varepsilon)$, endowed
with the dual norm $|u|_{-s} = \sup_{z \in V^s_\varepsilon,\, z \neq 0}(
\langle u , z \rangle_{V'_\varepsilon, V_\varepsilon} /{|z|_s})$.

Finally, let $B_\varepsilon$ be the bilinear form on $V_\varepsilon$
defined, for $(u_1,u_2) \in V_\varepsilon \times V_\varepsilon$, by
\begin{equation*}
\langle B_\varepsilon (u_1 , u_2),u_3\rangle_{V'_\varepsilon, V_\varepsilon} \,
=\, \int_{Q_\varepsilon} (u_1 \cdot \nabla)u_2 \cdot u_3 dx\,
\quad \forall u_3
\in V_\varepsilon~.
\end{equation*}

In order to simplify, we assume, in the whole paper (except in 
Theorem \ref{theoreme1bis}),  that the data $u_0$
and $f$ satisfy the conditions
\begin{equation}
\label{Hgener}
u_0 \in V^s_\varepsilon\, \quad {\rm for}~ {\rm some}\,s~,~ 0\,
\leq s \leq \, 1~, \quad f \in L^{\infty}(0, + \infty; H_\varepsilon)~.
\end{equation}
In Theorem \ref{theoreme1bis}, we shall suppose that $f \in 
L^2(0, + \infty; H_\varepsilon)$.
The Navier-Stokes equations, supplemented with the boundary conditions
(PP) or (PD) can then be written as a differential equation in
$V'_\varepsilon$:
\begin{equation}
\label{NSabs}
\begin{split}
&\partial_t u +\nu  A_\varepsilon u + B_\varepsilon (u,u) \, =\,
P_\varepsilon f~, \\
& u( \cdot, 0)\, =\, u_0~.
\end{split}
\end{equation}
Here $\partial_t u $ denotes the derivative (in the sense of distributions)
of $u$ with respect to $t$.

We now recall three classical existence
results of solutions to (\ref{NSabs}) (see \cite{CannoneLivre},
\cite{ConstantinFoias},
\cite{FujitaKato}, \cite{Ladyzhenskaya}, \cite{Leray},
 \cite{Lions},
\cite{Temam84}, \cite{VonWahl85}, \dots), which are valid if 
$f$ belongs to $L^{\infty}(0, + \infty; H_\varepsilon)$ or to 
$L^2(0, + \infty; H_\varepsilon)$:

\textbullet (P1)  For $u_0 \in H_\varepsilon$, there exists a solution $u$ of
(\ref{NSabs}) (not necessarily unique), such that
\begin{equation}
\label{faible}
u \in L^2_{loc}([0, + \infty); V_\varepsilon) \cap L^{\infty}(0,
+\infty; H_\varepsilon) \cap L^{4/3}_{loc}([0, +\infty); V'_\varepsilon)
\end{equation}
and, for all $0 \leq t\leq + \infty$,
\begin{equation}
\label{ienergie}
\|u(t)\|^2_{L^2}+ 2\nu \int_0^t \|\nabla u(s)\|^2_{L^2} ds \,
\leq \,
\|u_0\|^2_{L^2}+ 2 \int_0^t (f(s), u(s)) ds~.
\end{equation}
A solution $u$ of (\ref{NSabs}) satisfying (\ref{faible}) and
(\ref{ienergie}) is called a weak Leray solution.

\textbullet (P2) For $u_0 \in V_\varepsilon$, there exist a time
$T_\varepsilon = T_\varepsilon(Q_\varepsilon, \nu, u_0, P_\varepsilon
f)$ and a unique solution $u$ of (\ref{NSabs}),  such that
\begin{equation}
\label{fort}
u \in L^2_{loc}([0, T_\varepsilon;) V^2_\varepsilon) \cap L^{\infty}_{loc}([0,
T_\varepsilon); V_\varepsilon) \cap C^0([0,T_\varepsilon); V_\varepsilon)~.
\end{equation}
Such a solution is usually called a strong solution of (\ref{NSabs}).

\textbullet (P3) For $u_0 \in V^{1/2}_\varepsilon$, there exist a time
$T^*_\varepsilon = T^*_\varepsilon(Q_\varepsilon, \nu, u_0, P_\varepsilon
f)$ and a unique solution $u$ of (\ref{NSabs}), such that
\begin{equation}
\label{H1/2}
u \in L^2_{loc}([0, T^*_\varepsilon); V^{3/2}_\varepsilon) \cap
L^{\infty}_{loc}([0,T^*_\varepsilon); V^{1/2}_\varepsilon) \cap
C^0([0,T^*_\varepsilon); V^{1/2}_\varepsilon)~.
\end{equation}

Furthermore, using a classical small data argument like in \cite{RaugelSell93},
for instance, one shows that, if
\begin{equation}
\label{smalldata}
\|A^{1/4}_\varepsilon u_0\|_{L^2} + \sup_s \|A_\ep^{-1/4} P_\varepsilon
f(s)\|_{L^2}
\, \leq \, C \varepsilon^{1/2}~,
\end{equation}
where $C$ is independent of $\varepsilon$, then the solution $u$ of
(\ref{NSabs}) is global in time, that is, $T^*_\varepsilon =+ \infty$.

Here, we improve this global existence result as well
as those of \cite{Iftimie9999}, \cite{MoiseTemamZiane},
\cite{MontgomerySmith}, \cite{RaugelSell93}, \cite{RaugelSell94} and
\cite{TemamZiane96}. Before giving the precise statements, 
we need to define the mean value operator $M$ in the vertical 
direction:
\begin{equation}
\label{meanthin}
(Mf)(x_1,x_2)\, =\, \frac{1}{\varepsilon} \int_{0}^{\varepsilon}
f(x_1,x_2,s) ds\, \quad \forall f \in L^2(Q_\varepsilon)~.
\end{equation}
We extend this operator $M\in {\mathcal L}(L^2(Q_\varepsilon);
L^2(Q_\varepsilon))$ to an operator in ${\mathcal
L}((L^2(Q_\varepsilon))^3;
(L^2(Q_\varepsilon))^3)$ by setting $Mu =(Mu_1,Mu_2,Mu_3)$, for any
vector $u \in (L^2(Q_\varepsilon))^3$. Clearly, $M$ and $I-M$ are orthogonal
projections in $L^2(Q_\varepsilon)$ and $(L^2(Q_\varepsilon))^3$ and
commute with the derivations $D_i$, for $i=1,2,3$. Moreover, $MH_\varepsilon
\subset H_\varepsilon$. Using these properties and the fact that
$P_\varepsilon$ is an orthogonal projection onto $H_\varepsilon$, one
shows that
\begin{equation}
\label{MP}
MP_\varepsilon u \, =\, P_\varepsilon M u ~, \quad \forall u \in
(L^2(Q_\varepsilon)^3~,
\end{equation}
which implies that
\begin{equation}
\label{MA}
MA_\varepsilon u \, =\, A_\varepsilon M u ~, \quad \forall u \in
D(A_\varepsilon)~.
\end{equation}
One directly deduces from (\ref{MA}) that $M$ also commutes with the
operator $A^s_\varepsilon$, for $s \geq0$. 

The Navier-Stokes equations can now be rewritten as a system of
equations for $v \equiv Mu$ and $w \equiv (I-M)u$
\begin{equation}
\label{NSabsv}
\begin{split}
&\partial_t v +\nu  A_\varepsilon v + MB_\varepsilon (v,v)
 + MB_\varepsilon (w,w)\, =\,
MP_\varepsilon f~, \\
& v( \cdot, 0)\, =\, Mu_0~,
\end{split}
\end{equation}
and
\begin{equation}
\label{NSabsw}
\begin{split}
&\partial_t w +\nu  A_\varepsilon w + (I-M)(B_\varepsilon (v,w)
+ B_\varepsilon (w,v) + B_\varepsilon (w,w))\, =\,
(I-M)P_\varepsilon f~, \\
& w( \cdot, 0)\, =\, (I-M)u_0~.
\end{split}
\end{equation}

In the case of the (PD) and (PP) boundary conditions, we show
the following results.

\begin{theoreme}
\label{theoreme1}
Let $\varepsilon_0 > 0$ be fixed. For any  nonnegative numbers
$\alpha$, $\beta$, $\gamma$, $\delta$, with
$0 \leq \beta <1$, $0 \leq \gamma < 1/2$, $0 < \delta < 1/2$, 
there exists a positive constant $K_\ast = K_\ast
(\alpha,\beta,\gamma,\delta)$ such that, for $0
< \varepsilon \leq \varepsilon_0$, if the initial data $(Mu_0,(I-M)u_0)
\in H_\varepsilon \times V^{1/2}_\varepsilon$ and the forcing
term $f \in L^\infty(0,\infty;(L^2(Q_\varepsilon))^3)$ satisfy
\begin{equation}
  \label{10}
  \begin{split}
  &\nl{Mu_0}{2}\, \leq \,  K_\ast \varepsilon^{ - \alpha +1/2}~,\quad
  |(I-M)u_0|_{1/2}\,  \leq \,  K_\ast~,\\
  &\sup_t \nl{MP_\varepsilon f(t)}{2}\, \leq \,
 K_\ast \varepsilon^{-\beta +1/2 }~,\quad
  \sup_t \nl{(I-M)P_\varepsilon f(t)}{2}\, \leq \,
  K_\ast \varepsilon^{-1/2} |\ln \varepsilon|^\gamma~,
  \end{split}
\end{equation}
and the additional condition
\begin{multline}
  \label{10bis}
\left( |(I-M)u_0|_{1/2}^2 +
  \varepsilon^2 \sup_t \|(I-M)P_\varepsilon f(t)\|^2_{L^2}
 \exp K_\ast^{-1} (\varepsilon^{-1/2} 
  \sup_t \|MP_\varepsilon f(t)\|_{L^2})^{2(1+\delta)} \right) \\  
\times \exp K_\ast^{-1} (\varepsilon^{-1/2} \|Mu_0\|_{L^2})^{2(1+\delta)}  \,
  \leq \, K_\ast~,
\end{multline}
then the equations (\ref{NSabs}) admit a global solution $u
\in C^0([0,+\infty);H_\varepsilon)
\cap L^2_{loc}([0,+\infty);V_\varepsilon)
\cap H^1_{loc}([0, +\infty);V'_\varepsilon)$,
which is unique in the class of weak Leray solutions.
Moreover, $(I-M)u\in L^\infty(0,\infty;V^{1/2}_\varepsilon)\cap
L^2_{loc}([0,\infty);V^{3/2}_\varepsilon)$ and 
the estimates (\ref{4wfin}) and (\ref{4vfin}) hold, 
for $t \geq 0$.
\end{theoreme}

\begin{remarques} \label{Remarks1}
\noindent{\bf i)} In the particular case $\alpha = \beta =0$, the condition
(\ref{10bis}) always holds, provided the constant $K_\ast$ is 
small enough. The proof of Theorem \ref{theoreme1} shows that 
the condition (\ref{10bis}) can be replaced by a weaker hypothesis 
(see (\ref{4wfin})). 
In particular, if $\gamma=0$, (\ref{10bis}) can be replaced by the 
following weaker condition
\begin{multline*}
 \left( |(I-M)u_0|_{1/2}^2 +
  \varepsilon^3 \sup_t \|(I-M)P_\varepsilon f(t)\|^2_{L^2}
 \exp K_\ast^{-1} (\varepsilon^{-1/2} 
  \sup_t \|MP_\varepsilon f(t)\|_{L^2})^{2(1+\delta)} \right) \\  
\times \exp K_\ast^{-1} (\varepsilon^{-1/2} \|Mu_0\|_{L^2})^{2(1+\delta)}  \,
  \leq \, K_\ast~.
\end{multline*}

\noindent{\bf ii)}  In the case of periodic boundary conditions, 
we can set $\delta=0$, $\beta =1$, $\gamma =1/2$ 
in the hypothesis (\ref{10}). Moreover, the limitation on 
$\|Mu_0\|_{L^2}$ disappears and the condition (\ref{10bis}) 
simply writes
\begin{multline}
  \label{10bisbis}
 \left( |(I-M)u_0|_{1/2}^2 + \varepsilon^2 \sup_t \|(I-M)P_\varepsilon f(t)\|^2_{L^2}
  \exp K_\ast^{-1} \varepsilon^{-1} 
  \sup_t \|MP_\varepsilon f(t)\|^2_{L^2}  \right)\\
\times \exp K_\ast^{-1} \varepsilon^{-1} \|Mu_0\|^2_{L^2}
  \, \leq \, K_\ast~.
\end{multline}
This improvement will be explained in Remark \ref{Remark41}.

 \noindent{\bf iii)}  Applying the Poincar\'{e} inequality (\ref{Poincarew})
to the term $(I-M)u_0$, we easily see that the above theorem still
holds if, in the conditions (\ref{10}), (\ref{10bis}),
$|(I-M)u_0|_{1/2}$ is replaced by $\varepsilon^{1/2} |(I-M)u_0|_1$.
\end{remarques}

The above theorem has already been proved in \cite{Iftimie9999}, in
the frame of anisotropic spaces and Littlewood-Paley theory, in the
particular case of periodic boundary conditions and vanishing forcing
term $f$. 

\begin{remarque}\label{Remark4}
We also improve the results of \cite{Avrin}
in the case of homogeneous Dirichlet boundary conditions, by
requiring less regularity on the initial data $u_0$.
In this case, we
introduce the Laplace operator $\Delta_{dd}
=\Delta$, with domain $D(-\Delta_{dd}) = \{g \in H^1_0(Q_\varepsilon);
\Delta g \in L^2(Q_\varepsilon) \}$, where $H^1_0(Q_\varepsilon)$
is the closure in $H^1(Q_\varepsilon)$ of those smooth functions that
have compact support in $Q_\varepsilon$.
For $0 \leq s \leq 2$, we define the
space $X^s \equiv
D((-\Delta_{dd})^{s/2})$ equipped with the norm
$\|\cdot\|_s \equiv \|(-\Delta_{dd})^{s/2} \cdot\|_{L^2}$. If
$ H_\varepsilon =\left\{u \in (L^2(Q_\varepsilon))^3; \,{\rm
 div}u=0; \,  u \cdot \nu_\varepsilon  = 0 \, {\rm on}\,
 \partial Q_\varepsilon \right\}$,
$V_\varepsilon  =\left\{u \in H_\varepsilon; \, u = 0 \, {\rm on}\,
 \partial Q_\varepsilon \right\}$, where $\nu_\varepsilon $ is the outer
 normal to the boundary $\partial Q_\varepsilon $, 
 we define the corresponding Stokes operator $A_{\varepsilon}$ 
 with domain $D(A_\varepsilon) \equiv \left\{ u \in V_\varepsilon;
A_\varepsilon \in H_\varepsilon \right\}$. From
 \cite{Dauge}, it follows that $D(A_\varepsilon) =
 (H^2(Q_\varepsilon))^3 \cap V_\varepsilon$. Arguing as in
 \cite{TemamZiane96}, one shows that the Cattabriga-Solonnikov
 inequality (\ref {Catta}) holds uniformly in $\varepsilon$ and that
 the inequalities (\ref {Cattas}) are still true. We remark that the
 first eigenvalue of $-\Delta_{dd}$ (respectively $A_\varepsilon$) is
 of order $\varepsilon^{-2}$, which implies that the Poincar\'{e} 
 inequalities (\ref{Poincarew}) and (\ref{Poincarewbis}) hold, with 
 $(I-M)f$ replaced by $f$.
 Hence, the decomposition $u=Mu + (I-M)u$ is of no use. 
Replacing simply $w$ by $u$ and $v$ by $0$ in
 the proof of Theorem \ref{theoreme1}, one
 shows that there exists a positive
 constant $K$ such that, for $0< \varepsilon\leq\varepsilon_0$,
if $u_0 \in V^{1/2}_\varepsilon$ and 
$f\in L^\infty(0,\infty;L^2(Q_\varepsilon))$ satisfy
\begin{equation}
  \label{DD10}
  |u_0|_{1/2}\,  \leq \, K~,\quad
  \sup_t \nl{P_\varepsilon f(t)}{2} \, \leq \, K
  \varepsilon^{-3/2}~,
\end{equation}
then the equations (\ref{NSabs}) admit a global solution $u \in C^0([0,+\infty);H_\varepsilon)
\cap L^\infty(0,\infty;V^{1/2}_\varepsilon) \cap
L^2_{loc}([0,\infty);V^{3/2}_\varepsilon)$,which is
unique in the class of weak Leray solutions. Moreover, there exists
a positive constant $C$ independent of $\varepsilon$, such that,
for $t\geq 0$,
\begin{equation}
\label{1estu}
|u(t)|^2_{1/2} \, \leq \, \exp(-C\varepsilon^{-2}t) |u_0|^2_{1/2} +
C\varepsilon^3 \sup_t \|(I-M) P_{\varepsilon}f(t)\|^2_{L^2}~.
\end{equation}
\end{remarque}

\begin{remarque}\label{Remark5}
As in \cite{TemamZiane96}, if
$\Omega = (0,l_1) \times (0,l_2)$, we can consider the
Navier-Stokes equations (\ref{NS}), supplemented with the (DP)
boundary conditions, that is, homogeneous Dirichlet boundary
conditions on $\Gamma_v = (\Omega \times \{x_3= 0\}) \cup
(\Omega \times \{x_3= \varepsilon\})$ and periodic
conditions in the variables $x_1$, $x_2$. 
As before, one defines the
corresponding spaces $X^s$, $H_\varepsilon$, $V_\varepsilon$ and
the corresponding Stokes operator $A_\varepsilon$. The
inequalities (\ref{Catta}) and (\ref{Cattas}) still hold. Thus, like
in Remark \ref{Remark4}, ones proves that there exists a positive
constant $K$ such that, for $0< \varepsilon\leq\varepsilon_0$,
if $u_0 \in V^{1/2}_\varepsilon$ and 
$f\in L^\infty(0,\infty;L^2(Q_\varepsilon))$ satisfy
the conditions (\ref{DD10}),
then the equations (\ref{NSabs}) admit a global solution $u \in C^0([0,+\infty);H_\varepsilon)
\cap L^\infty(0,\infty;V^{1/2}_\varepsilon) \cap
L^2_{loc}([0,\infty);V^{3/2}_\varepsilon)$, which is
unique in the class of weak Leray solutions,
and the estimate (\ref{1estu}) holds.
\end{remarque}

We now assume that the forcing term $Pf$ belongs to $L^2(0,\infty;
(L^2(Q_\varepsilon))^3)$, which is a rather strong requirement.
But, in this case, we can remove every smallness assumption on
the data $Mu_0$ and $MPf(t)$, provided the data $w_0$ and
$(I-MP)f(t)$ are small enough. 
\begin{theoreme}
\label{theoreme1bis}
Let $\varepsilon_0 > 0$ be fixed. For any  positive number
$\delta$, $0 < \delta <1/2$,
there exists a positive constant $\tilde{K} = \tilde{K}
(\delta)$ such that, for $0 < \varepsilon \leq \varepsilon_0$, 
if the initial data $(Mu_0,(I-M)u_0)
\in H_\varepsilon \times V^{1/2}_\varepsilon$ and the forcing
term $f \in L^2(0,\infty; (L^2(Q_\varepsilon))^3)$ satisfy
\begin{multline}
  \label{10ter}
  \left( |(I-M)u_0|_{1/2}^2 +
  \varepsilon \int_{0}^{+\infty}
  \|(I-M)P_\varepsilon f(\tau)\|^2_{L^2} d\tau\right) \\
  \exp \tilde{K}^{-1}
  \left( \varepsilon^{-1} \|Mu_0\|^2_{L^2} + \varepsilon^{-1}
 \int_{0}^{+\infty} \|MP_\varepsilon f(\tau)\|^2_{L^2} 
 d\tau \right)^{1+\delta}\, \leq \, \tilde{K}~,
\end{multline}
then there exists a global solution $u
\in C^0([0,+\infty);H_\varepsilon)
\cap L^2_{loc}([0,+\infty);V_\varepsilon)
\cap H^1_{loc}([0, +\infty);V'_\varepsilon)$
of (\ref{NSabs}) which is unique in the class of weak Leray solutions.
Moreover, $(I-M)u\in L^\infty(0,\infty;V^{1/2}_\varepsilon)\cap
L^2_{loc}([0,\infty);V^{3/2}_\varepsilon)$.
\end{theoreme}
\vskip 2mm
If, for instance, in Theorem \ref{theoreme1}, we want to choose $Mu_0$ of
order $\varepsilon^{\theta}$, for $\theta < 1/2$, we need to assume
that $(I-M)u_0$ and
$(I-M)P_\varepsilon f$ are exponentially small functions of
$\varepsilon$. However, in the case of the (PP) boundary conditions,
these drastic restrictions become much milder.
In the theorem below, we split the vector field $v \equiv Mu$
into two parts
\begin{equation}
\label{decoupe}
Mu \, = \, M\tilde{u} + M(u_3)\, \equiv \, (Mu_1,Mu_2,0) + (0,0, Mu_3)~,
\end{equation}
and set $\tilde{v} = M\tilde{u}$. In the proof, we use the 
conservation of enstrophy for the vector field $\tilde{v}$.

\begin{theoreme} \label{theoreme2}
Let $\varepsilon_0>0$ be fixed. There exist positive
constants $k_1$, $k_2$, $k_3$, $k_4$ and $k_5$ such that, for
$0< \varepsilon\leq \varepsilon_{0}$, if the initial data
 $(Mu_0, (I-M)u_0) \in V_p \times V^{1/2}_p$ and the forcing term $f\in
L^\infty(0,\infty;(L^2(Q_\varepsilon))^3)$ satisfy
\begin{equation}
  \label{H1th}
 \begin{split}
  &|Mu_0|_1 \, \leq \, k_1 \varepsilon^{-1/2}, \quad |(I-M)u_0|_{1/2}
  \,\leq \, k_2\\
  &\sup_t \nl{MP_\varepsilon f(t)}{2}\leq k_3 \varepsilon^{-1/2},
 \quad \sup_t \nl{(I-M)P_\varepsilon f(t)}{2}\, \leq \, k_4
 \varepsilon^{- 1} ~,
\end{split}
\end{equation}
and the additional condition
\begin{multline}
\label{H2th}
\mathcal{A}_0 \, \equiv\, \Big( |M\tilde{u}_0|_1 +
\sup_t \nl{M\widetilde{P_\varepsilon f}(t)}{2}
+ \varepsilon^{-1/2} |(I-M)u_{0}|_{1/2}^2 +
 \varepsilon^{3/2} \sup_t \|(I-M)P_\varepsilon f(t)\|^2_{L^2} \Big) \\
\times\Big( \nl{M(u_{03})}{2} + \sup_t |M(P_\varepsilon f)_3(t)|_{-1} \Big)
\, \leq \, k_5~,
\end{multline}
then the equations (\ref{NSabs}) admit a global solution $u(t) \in C^0([0,\infty); V^{1/2}_p) \cap
L^{\infty}(0,\infty; V^{1/2}_p) \cap L^2_{loc}([0,\infty); V^{3/2}_p)$,
which is unique in the class of weak Leray solutions.
Moreover, $Mu$ belongs to the space $C^0([0,\infty); V_p) \cap
L^\infty(0,\infty;V_p)\cap L^2_{loc}([0,\infty);V^2_p)$
and the estimates (\ref{est1w}) and (\ref{contra}) hold, for
every $t\geq 0$.
\end{theoreme}

\begin{remarque}\label{RemarkFP}
Similar existence results hold, if one considers the Navier-Stokes 
equations (\ref{NS}), supplemented with the (FP) boundary conditions, 
that is, with the free boundary condition 
\begin{equation}
\label{FP3}
u_3(x_1,x_2,x_3)\, = \, 0~, \quad 
\partial_{x_3}u_j(x_1,x_2,x_3)\, = \, 0~, \quad j=1,2~, 
\quad x_3\,=\, 0,\varepsilon~,
\end{equation}
and periodic conditions in the variables $x_1$, $x_2$. As before, one 
defines the corresponding spaces $H_\varepsilon$, $V_\varepsilon$ and 
the Stokes operator $A_\varepsilon$. In the proofs of Section 2, one 
also needs to define the spaces $X^s$, which are now different for 
$u_j$, $j=1,2$ and $u_3$. Since 
$u_3(x_1,x_2,0)=u_3(x_1,x_2,\varepsilon)=0$, one introduces the 
following mean value operator  $M_{FP}$ on $H_\varepsilon$,
\begin{equation*}
M_{FP}u \, = \, (Mu_1,Mu_2,0)~, \quad \forall u \in H_\varepsilon~.
\end{equation*}
Then, one easily checks that the theorems \ref{theoreme1},
\ref{theoreme1bis} and \ref{theoreme2} are still true, if the operator 
$M$ is replaced by the corresponding operator $M_{FP}$. Remark that, 
since $M_{FP}(0,0, u_{03})=M_{FP}(0,0, (P_\varepsilon f)_3)=0$, the 
additional condition (\ref{H2th}) disappears. The proof of Theorem 
\ref{theoreme2} in the (FP) case is actually much simpler than in the periodic case, 
because the term $v_3$ is zero. Also in the (FP) case, Theorem \ref{theoreme2} 
improves the corresponding result of \cite{TemamZiane96}. 
\end{remarque}

In Theorem \ref{theoreme3} of Section 5 , we shall give another global
existence and uniqueness result, involving the $L^p$-norm of $Mu_{03}$.
As a direct consequence of Theorem \ref{theoreme2} , we obtain
the following simple corollary:

\begin{corollaire}\label{corollaire1}
Let $\varepsilon_0>0$ be fixed. There exists a positive constant $k_0$,
 such that, for $0< \varepsilon\leq \varepsilon_{0}$, for $0
 \leq \beta \leq 1/2$, if the initial data
 $(Mu_0, (I-M)u_0) \in V_p \times V^{1/2}_p$ and the forcing term $f\in
L^\infty(0,\infty;(L^2(Q_\varepsilon))^3)$ satisfy
\begin{equation}
\begin{split}
\label{donnees1}
& |M\tilde{u}_0|_1 + \sup_t \nl{M\widetilde{P_\varepsilon f}(t)}{2} \, \leq
\, k_0 \varepsilon^{-\beta}\, , \quad
\sup_t \nl{(I-M)P_\varepsilon f(t)}{2}\, \leq \, k_0 \varepsilon^{-3/4
-\beta/2}~, \\
& |Mu_{03}|_1 + \sup_t \nl{MP_\varepsilon f_3(t)}{2} \, \leq \,
k_0\varepsilon^{-1/2}\, , \quad
 \nl{Mu_{03}}{2}
+ \sup_t |MP_\varepsilon f_3(t)|_{-1}\, \leq \, k_0 \varepsilon^\beta~,
 \end{split}
 \end{equation}
and
\begin{equation}
\label{donnees2}
 |(I-M)u_0|_{1/2} \, \leq \, k_0 \varepsilon^{1/4 -\beta/2}~,
 \end{equation}
 then the equations (\ref{NSabs}) admit 
 a global solution $u(t) \in C^0([0,\infty); V^{1/2}_p) \cap
L^{\infty}(0,\infty; V^{1/2}_p) \cap L^2_{loc}([0,\infty); V^{3/2}_p)$,
which is unique in the class of weak Leray solutions.
Moreover, $Mu$ belongs to the space $C^0([0,\infty); V_p) \cap
L^\infty(0,\infty;V_p)\cap L^2_{loc}([0,\infty);V^2_p)$
and the estimates (\ref{est1w}) and (\ref{contra}) hold, for
every $t\geq 0$.
\end{corollaire}

Applying the Poincar\'{e} inequality (\ref{Poincarewbis}) to $(I-M)u_0$,
we at once get the following global existence result:

\begin{corollaire}\label{corollaire2}
Let $\varepsilon_0>0$ be fixed. There exists a positive constant $k_0$,
 such that, for $0< \varepsilon\leq \varepsilon_{0}$, for $0
 \leq \beta \leq 1/2$, if the initial data
 $u_0 \in V_p$ and the forcing term $f\in
L^\infty(0,\infty;(L^2(Q_\varepsilon))^3)$ satisfy the conditions
(\ref{donnees1})
and
\begin{equation}
\label{donnees2bis}
 |(I-M)u_0|_1 \, \leq \, k_0 \varepsilon^{-1/4 -\beta/2}~,
 \end{equation}
then the equations (\ref{NSabs}) have a unique
 global strong solution $u(t) \in C^0([0,\infty);V_p)\cap
 L^2_{loc}([0,\infty);V^2_p)$.
\end{corollaire}

\begin{remarque}\label{Remark6}
In \cite{RaugelSell94}, it has been proved, in the (PP) case, that
there exists $\varepsilon_1 >0$ such that, for $0< \varepsilon \leq \varepsilon_1$, 
the equations (\ref{NSabs}) admit a unique global strong solution 
$u \in C^0([0,\infty);V_\varepsilon)$, if the data satisfy the 
following conditions, where $\delta$ is a small positive constant,
\begin{equation}
\begin{split}
\label{RSell1}
& |Mu_0|_1 + \sup_t \|MP_\varepsilon f(t)\|_{L^2} \, \leq \, C
\varepsilon^{\delta +7/24}~, \\
& |(I - M)u_0|_1 \, \leq \, C \varepsilon^{\delta  - 5/48}\, , \quad
\sup_t \|(I - M)P_\varepsilon f(t)\|_{L^2} \, \leq \, C
\varepsilon^{\delta- 1/2 }~, \\
\end{split}
\end{equation}
or
\begin{equation}
\begin{split}
\label{RSell2}
& |Mu_0|_1 \, \leq \, C \varepsilon^{\delta -1/32}\, , \quad
\sup_t \|MP_\varepsilon f(t)\|_{L^2} \, \leq \, C \varepsilon^{\delta
-1/16 }~, \\
& |(I - M)u_0|_1 \, \leq \, C \varepsilon^{\delta  - 1/8}\, , \quad
\sup_t \|(I - M)P_\varepsilon f(t)\|_{L^2} \, \leq \, C
\varepsilon^{\delta - 1/2 }~, \\
& \nl{Mu_{03}}{2}
+ \sup_t \|MP_\varepsilon f_3(t)\|_{L^2} \, \leq \, k_0 \varepsilon~.
\end{split}
\end{equation}
In \cite{MoiseTemamZiane}, Moise, Temam and Ziane have shown that, in
the (PP) case,  there exists
$\varepsilon_1 >0$ such that, for $0< \varepsilon \leq \varepsilon_1$,
the equations (\ref{NSabs}) admit a unique global strong solution 
$u \in C^0([0,\infty);V_\varepsilon)$, if the data satisfy the 
following conditions, where $\delta$ is a small positive constant,
\begin{equation}
\begin{split}
\label{MTZ}
& |Mu_0|_1 + \sup_t \|MP_\varepsilon f(t)\|_{L^2} \, \leq \, C
\varepsilon^{\delta +1/6 }~, \\
& |(I - M)u_0|_1 + \sup_t \|(I - M)P_\varepsilon f(t)\|_{L^2} \, \leq \, C
\varepsilon^{\delta- 1/6 }~. 
\end{split}
\end{equation}
Choosing $\beta = 0$ in Corollary \ref{corollaire2}, one at once sees
that the conditions (\ref{donnees1})
and (\ref{donnees2bis}) allow larger data than the hypotheses
(\ref{RSell1}), (\ref{RSell2}) or (\ref{MTZ}).
Finally, Corollary \ref{corollaire2} improves as well the results of
\cite{MontgomerySmith}, where global existence and uniqueness are proved
under the
assumption $|u_0|_1 + \sup_t \|P_\varepsilon f(t)\|_{L^2} \leq  C $,
for some constant $C$.
\end{remarque}

An outline of the paper is as follows. In order to estimate the quadratic
term in (\ref{NSabs}), we prove some auxiliary inequalities in Section 2.
Section 3 is devoted to a uniqueness result. In
Section 4, we give the proofs of the theorems \ref{theoreme1} and
\ref{theoreme1bis}. Section 5 contains
the proofs of the theorems \ref{theoreme2} and \ref{theoreme3}.

In the sequel, we
shall write $A$ and $P$ for the operator $A_\varepsilon$ and the
projection $P_\varepsilon$. The
constants $K, K_1,\ldots$ and $C, C_1,\ldots$ will always denote
positive constants, that are independent of $\varepsilon$. We recall 
that we denote the spaces $\dot{H}^s_p$ or $H^s_d$ by $X^s$, when no
distinction concerning the boundary conditions is necessary.

\section{Auxiliary estimates}

In (\ref{meanthin}) we have introduced the mean value operator $M \in 
\mathcal{L}(L^2(Q_\varepsilon); L^2(Q_\varepsilon))$ and extended it 
to an operator $M \in \mathcal{L}((L^2(Q_\varepsilon))^3; 
(L^2(Q_\varepsilon))^3)$, by setting $Mu=(Mu_1,Mu_2,Mu_3)$. This 
operator $M$ allows to decompose every function $f \in 
L^2(Q_\varepsilon)$ into $f=Mf + (I-M)f$, where $Mf$ is a function of 
$x_1$ and $x_2$ only and $(I-M)f$ satisfies the following Poincar\'{e} 
inequality
\begin{equation}
\label{Poincarew}
\|(I-M)f\|_{L^2}\, \leq \, \tilde{K}_0 \|(I-M)f\|_{H^s}\, \leq \,
K_0 \varepsilon^s \|(I-M)f\|_s\, ,
\quad \forall f \in X^s, \quad
0\, \leq \, s \, \leq \, 2~,
\end{equation}
where $\tilde{K}_0$, $K_0$ are independent of $s$, $f$ and
$\varepsilon$ (see \cite{HaleRaugel9292},
\cite{HaleRaugel929292}, for instance). We notice that the
constant $K_0$ in the inequality (\ref{Poincarew}) can be chosen so that
\begin{equation}
\label{Poincarewbis}
\|(I-M)u\|_{L^2}\, \leq \, K_0 \varepsilon^s |(I-M)u|_s\, ,
\quad \forall u \in V^s_{\varepsilon}, \quad
0\, \leq \, s \, \leq \, 2~.
\end{equation}
These inequalities will be often used  below.

We shall also need the following classical Poincar\'{e} inequalities,
for $0 \leq s \leq 2$,
\begin{equation}
\label{Poincare}
\|u\|_{L^2} \, \leq \, \mu^s_0 \|u\|_s~, \quad \forall u \in X^s~,
\end{equation}
and
\begin{equation}
\label{Poincarebis}
\|u\|_{L^2} \, \leq \, \mu^s_0 |u|_s~, \quad \forall u \in 
V^s_\varepsilon~,
\end{equation}
where $\mu_0$ is a positive constant depending only on $\Omega$. 

We denote by $L^{q,q'}(Q_{\varepsilon})=
L^q(\Omega;L^{q'}(0,\varepsilon))$
or simply $L^{q,q'}$ the space of (classes of) functions $u$ such that
$\nl{u}{q,q'}=\|\|u\|_{L^{q'}_{x_3}(0,\varepsilon)}\|_{L^q_{x'}(\Omega)}$ is
finite, where $x'=(x_1,x_2)$. Of course, $L^{q,q}$ is the
usual space $L^q(Q_\varepsilon)$ and the norm
$\nl{u}{q,q}$ is denoted by $\|u\|_{L^q}$.

The following property of a divergence-free vector field will also be
frequently used:
\begin{equation}
\label{gradient}
\|\nabla u\|^2_{L^2(Q_\ep)}
= \sum_i\|\nabla u_i\|^2_{L^2(Q_\ep)}
= -\int _{Q_\ep}u \cdot \Delta u
= -\int _{Q_\ep}u \cdot A u
=\|A^{1/2} u\|^2_{L^2(Q_\ep)}~.
\end{equation}
\begin{lemme}\label{lema1}
Let $\varepsilon_0 > 0$ be fixed. There exists a positive constant
  $K_1$ so that, for $0 < \varepsilon \leq \varepsilon_0$,
if $w_i \in X^{s_i}$ are three functions satisfying $Mw_i=0$,
$0\leq s_i<3/2$, for $i=1,2,3$, and $s_1+s_2+s_3=3/2$, then 
  \begin{equation}
    \label{unu}
    \left|\int_{Q_\ep}w_1(x) w_2(x) w_3(x)\,dx\right|\, \leq \, K_1
    \|w_1\|_{s_1} \|w_2\|_{s_2} \|w_3\|_{s_3}.
  \end{equation}
 Furthermore, there exists a positive constant $K_2$ such that, 
 for $0 < \varepsilon \leq \varepsilon_0$,
 if $v_1 \in X^{\tilde{s}_1}$, $0 \leq \tilde{s}_1 <1$
is a function independent of $x_3$, $0 \leq s_i<1$, for $i=2,3$ and
 $\tilde{s}_1+s_2+s_3=1$, then 
  \begin{equation}
    \label{doi}
    \left|\int_{Q_\ep}v_1(x) w_2(x) w_3(x)\,dx\right| \, \leq \, 
     K_2 \ep^{-1/2}
    \|v_1\|_{\tilde{s}_1} \|w_2\|_{s_2} \|w_3\|_{s_3}.
  \end{equation}
\end{lemme}

\begin{remarque}
\label{remarque}
It will be clear from the proof below that, if we omit the 
dependence on $\ep$,
Lemma \ref{lema1} still holds for functions without vanishing 
mean in the thin direction.
\end{remarque}

Lemma \ref{lema1} is a consequence of the following result.
  
\begin{lemme}\label{lema2}
Let $\varepsilon_0 > 0$ be fixed.
 Assume that $0\leq s<3/2$ and $q,q'\in [2,\infty)$ are such that
    $\frac{2}{q}+\frac{1}{q'}=\frac{3}{2}-s$. Then the following
    embedding holds:
    \begin{equation*}
    X^s\hookrightarrow L^{q,q'}(Q_\varepsilon)~.
    \end{equation*}
 Moreover, there exists a positive constant $K_3$ 
such that , for $0 < \varepsilon \leq \varepsilon_0$,
 for any $w \in X^s$ satisfying $Mw = 0$,
    \begin{equation}
      \label{trei}
      \nl{w}{q,q'}\leq  K_3\|w\|_s.
    \end{equation}
  \end{lemme}
  
 \begin{proof}[Proof of Lemma \ref{lema1}]
Let us assume that Lemma  \ref{lema2} is proved. The particular 
case
$q=q'$ implies the embedding $X^s\hookrightarrow L^q(Q_\varepsilon)$
provided that $1/q=1/2-s/3$. Therefore, there exist three positive
constants $C_1$, $C_2$ and $C_3$ independent of $\varepsilon$
such that
\begin{equation*}
  \nl{w_i}{q_i}\, \leq \, C_i \|w_i\|_{s_i}~,\quad \forall i\in\{1,2,3\},
\end{equation*}
where $1/q_i=1/2-s_i/3$. Since $1/q_1+1/q_2+1/q_3=1$, 
H\"{o}lder's inequality gives
\begin{equation*}
  \left|\int_{Q_\ep}w_1(x) w_2(x) w_3(x)\,dx\right|
  \leq \nl{w_1}{q_1}\nl{w_2}{q_2}\nl{w_3}{q_3}
 \leq   C_1 C_2 C_3 \|w_1\|_{s_1} \|w_2\|_{s_2} \|w_3\|_{s_3}~,
\end{equation*}
which implies (\ref{unu}) with $K_1=C_1 C_2 C_3$.

Now we prove the inequality (\ref{doi}). As in the introduction, we define
the usual Hilbert spaces $H^s(\Omega)$ by interpolation, when
$s \geq 0$ is not an integer. Remarking that, for all $v \in
H^j(\Omega)$, $j \in {\mathbb N }$, $\|v\|_{H^j(\Omega)} =
\varepsilon^{-1/2}\|v\|_{H^j(Q_\varepsilon)}$, we deduce, by
interpolation, that, for $s \geq 0$,
\begin{equation}
\label{21auxil}
\|v\|_{H^s(\Omega)} \, \leq \, \varepsilon^{-1/2}
\|v\|_{H^s(Q_\varepsilon)}~, \quad \forall v \in H^s(\Omega)~.
\end{equation}
Due to the two-dimensional Sobolev embedding 
$H^{\tilde{s}_1}(\Omega) \hookrightarrow
L^{\tilde{q}_1}(\Omega)$, where $1/\tilde{q}_1 = (1-\tilde{s}_1)/2$,
and to the estimates (\ref{inegnorms}) and (\ref{21auxil}), we obtain
\begin{equation*}
  \|v_1\|_{L^{\tilde{q}_1}(\Omega)}\, \leq \,
  \tilde{C}  \|v_1\|_{H^{\tilde{s}_1}(\Omega)}\, \leq \,
\tilde{C} \varepsilon^{-1/2} \|v_1\|_{H^{\tilde{s}_1}(Q_\varepsilon)}
\, \leq \, \tilde{C}_1 \varepsilon^{-1/2} \|v_1\|_{\tilde{s}_1}~,
\end{equation*}
where $\tilde{C}_1 = c_1 \tilde{C}$ is a positive constant independent
of $\varepsilon$. On the other hand, one can apply
Lemma \ref{lema2} with $q'=2$ to get the existence of two constants
$\tilde{C}_2$ and $\tilde{C}_3$ independent of $\ep$ such that, for 
$i=2,3$,
\begin{equation*}
  \nl{w_i}{\tilde{q}_i,2}\, \leq \, \tilde{C}_i\|w_i\|_{s_i}~,
\end{equation*}
where $1/\tilde{q}_i=(1-s_i)/2$ for $i=2,3$. H\"older's
inequality adapted to the case of anisotropic spaces and the equality
$1/\tilde{q}_1+1/\tilde{q}_2+1/\tilde{q}_3=1$ yield
\begin{align*}
  \left|\int_{Q_\ep}v_1(x) w_2(x) w_3(x)\,dx\right|
  &\leq\|v_1\|_{L^{\tilde{q}_1}(\Omega)}\nl{w_2}{\tilde{q}_2,2}
  \nl{w_3}{\tilde{q}_3,2}\\
  &\leq \tilde{C}_1 \tilde{C}_2\tilde{C}_3\ep^{-1/2}
  \|v_1\|_{\tilde{s}_1} \|w_2\|_{s_2} \|w_3\|_{s_3},
\end{align*}
whence the inequality (\ref{doi}) with $K_2=\tilde{C}_1\tilde{C}_2
\tilde{C}_3$. The proof is completed.
\end{proof}

\begin{proof}[Proof of Lemma \ref{lema2}]
Let $d \geq 0$. Like in the introduction, we can define the operator 
$d - \Delta_2=d -\partial^2_{x_1x_1}-\partial^2_{x_2x_2}$ 
on $\Omega$, supplemented either with homogeneous Dirichlet boundary 
conditions in the (PD) case or with periodic boundary conditions in 
the (PP) case. Let $(\lambda_k,\varphi_k)_{k\geq 0}$ be a sequence 
of eigenvalues and eigenfunctions of $-\Delta_2$, such that
$(\varphi_k)_{k\geq 0}$ forms an orthonormal basis in 
$L^2(\Omega)$ and that $0 \leq \lambda_0 < \lambda_1 \leq \cdots$.
For $0 \leq \sigma \leq 2$, the operator $(d - 
\Delta_2)^{\sigma/2}$ writes, for any $v \in 
D((d - \Delta_2)^{\sigma/2})$,
\begin{equation*}
(d - \Delta_2)^{\sigma/2} v(x')\, = \, \sum_{k\geq 0}
(d+\lambda_k)^{\sigma/2}v_{k} \varphi_k(x')~,
\end{equation*}
where $v=\sum_{k\geq 0} v_{k} \varphi_k(x')$. We notice that 
$(e^{2i\pi nx_3/\varepsilon} \varphi_k(x'))_{n\in\mathbb{Z}, k\geq 0}$ 
is an orthonormal basis in $L^2(Q_\varepsilon)$ and the operator 
$(d-\Delta)^{\sigma/2}$ on $Q_\varepsilon$ (where $\Delta =\Delta_d$ or 
$\Delta =\Delta_p$ according to the boundary conditions) writes, 
for any $u \in X^\sigma$,
\begin{equation*}
(d - \Delta)^{\sigma/2} u(x)\, = \, \varepsilon^{-1/2} 
\sum_{n\in\mathbb{Z}, k\geq 0}
(d+\lambda_k + (\frac{2 \pi n}{\varepsilon})^2)^{\sigma/2}
u_{nk} e^{2\pi inx_3/\varepsilon}\varphi_k(x')~,
\end{equation*}
where $u= \varepsilon^{-1/2} \sum_{n\in\mathbb{Z}, k\geq 0} u_{nk} 
e^{2\pi inx_3/\varepsilon}\varphi_k(x')$. 

Again, like in the introduction, for $0 \leq \sigma \leq 2$, 
we define the Hilbert spaces 
$H^\sigma_p(0,\varepsilon)$ and $H^\sigma_p(0,1)$
of periodic functions on $(0,\varepsilon)$ 
and $(0,1)$. Performing a change of variables from $(0,\varepsilon)$ 
to $(0,1)$ and using the Sobolev embedding in dimension 1, 
$H^{\sigma'}_p(0,1)\hookrightarrow L^{q'}(0,1)$, where $\sigma' =
1/2- 1/q'$, we obtain, for any $g(x_3) \equiv \varepsilon^{-1/2} 
\sum_{n\in\mathbb{Z}} g_n e^{2 \pi inx_3/\varepsilon}$
in $H^{\sigma'}_p(0,\varepsilon)$,
\begin{equation}
\label{2g1}
\begin{split}
\|g\|_{L^{q'}(0,\varepsilon)} &\, = \, \varepsilon^{1/q'} 
\|\varepsilon^{-1/2} \sum_{n\in\mathbb{Z}} 
g_n e^{2 \pi iny}\|_{L^{q'}(0,1)} \\
&\, \leq \, C_1 \varepsilon^{1/q'} (\|\varepsilon^{-1/2} \sum_{n\in\mathbb{Z}} 
g_n e^{2 \pi iny}\|_{L^{2}(0,1)}
+ \|(-\partial^2_{yy})^{\sigma'/2}\varepsilon^{-1/2} \sum_{n\in\mathbb{Z}} 
g_n e^{2 \pi iny}\|_{L^{2}(0,1)}) \\
&\, \leq \, C_2 \varepsilon^{1/q'-1/2} ( \|g\|_{L^2(0,\varepsilon)}+
\varepsilon^{\sigma'}
\|(-\partial^2_{x_3x_3})^{\sigma'/2}g\|_{L^{2}(0,\varepsilon)})~.
\end{split}
\end{equation}
But we have
\begin{equation}
\label{2g2}
\|g\|_{L^2(0,\varepsilon)}+ \varepsilon^{\sigma'}
\|(-\partial^2_{x_3x_3})^{\sigma'/2}g\|_{L^{2}(0,\varepsilon)} \, 
\leq \, C_3 
\| (1-\partial^2_{x_3x_3})^{\sigma'/2} g\|_{L^{2}(0,\varepsilon)}~.
\end{equation}
If $g \in 
H^{\sigma'}_p(0,\varepsilon)$ satisfies $Mg=0$,  then,  
due to the Poincar\'{e} inequality (\ref{Poincarew}), we improve the 
above inequality and obtain
\begin{equation}
\label{2g3}
\begin{split}
\|g\|_{L^2(0,\varepsilon)}+ \varepsilon^{\sigma'}
\|(-\partial^2_{x_3x_3})^{\sigma'/2}g\|_{L^{2}(0,\varepsilon)} &\, 
\leq \, (K_0 +1) \varepsilon^{\sigma'}
\| (-\partial^2_{x_3x_3})^{\sigma'/2} g\|_{L^{2}(0,\varepsilon)} \\
&\, \leq \, C_4 \varepsilon^{\sigma'} 
\| (1-\partial^2_{x_3x_3})^{\sigma'/2} g\|_{L^{2}(0,\varepsilon)}~.
\end{split}
\end{equation}
The estimates (\ref{2g1}), (\ref{2g2}) and (\ref{2g3}) imply that
\begin{equation}
\label{2g4}
\|g\|_{L^{q'}(0,\varepsilon)} \, \leq \, \varepsilon^{1/q' -1/2} C_2 
C_0(\varepsilon) 
\|(1- \partial^2_{x_3x_3})^{\sigma'/2}g\|_{L^{2}(0,\varepsilon)}~,
\end{equation}
 where $C_0(\varepsilon)=C_3$ in the general case and 
$C_0(\varepsilon)=C_4\varepsilon^{\sigma'}$, when $Mg=0$.  
Let now $u$ be a function in $X^s$ and $q,q'\in [2,\infty)$. If
$\sigma' = 1/2 - 1/q'$, we deduce from (\ref{2g4}) that
\begin{equation*}
\begin{split}
\|u\|_{L^{q,q'}} \, = \, \| \|u\|_{L^{q'}_{x_3}} \|_{L^{q}_{x'}} \, 
&\leq \, \varepsilon^{1/q' -1/2} C_2 C_0(\varepsilon) 
\| \, \| (1- \partial^2_{x_3x_3})^{\sigma'/2}
u\|_{L^{2}_{x_3}}\|_{L^{q}_{x'}} \\
&\, \leq \, \varepsilon^{1/q' -1/2} C_2 C_0(\varepsilon) 
\| \, \| (1- \partial^2_{x_3x_3})^{\sigma'/2}
u \|_{L^{q}_{x'}}\|_{L^{2}_{x_3}}~,
\end{split}
\end{equation*}
where we could interchange the order of integrations, since 
$q \geq 2$. Now, the 2-dimensional Sobolev embedding 
$D((1 - \Delta_2)^{\sigma/2})) \hookrightarrow L^q(\Omega)$ with
$1/q=(1-\sigma)/2$ implies that
\begin{equation}
\label{patru}
\|u\|_{L^{q,q'}} \, \leq \, \varepsilon^{1/q' -1/2} C_2 
C_0(\varepsilon) C_5 \|(1 -\Delta_2)^{\sigma/2}
  (1-\partial^2_{x_3x_3})^{\sigma'/2}u\|_{L^{2}}~.
\end{equation}
But, as $\sigma + \sigma'=s$,
\begin{align*}
\|(1 -\Delta_2)^{\frac{\sigma}{2}}
  (1-\partial^2_{x_3x_3})^{\frac{\sigma'}{2}}u\|^2_{L^{2}} 
  & \, =\,
 \varepsilon^{-1} \| \sum_{n\in\mathbb{Z}, k\geq 0}
(1 + \lambda_k)^{ \frac{\sigma}{2}} 
(1+ (\frac{2 \pi n}{\varepsilon})^2)^{\frac{\sigma'}{2}}
u_{nk} e^{2\pi in \frac{x_3}{\varepsilon}}\varphi_k(x')\|^2_{L^{2}}\\
& \, = \,  \sum_{n\in\mathbb{Z}, k\geq 0}
(1 + \lambda_k)^{\sigma} 
(1+ (\frac{2 \pi n}{\varepsilon})^2)^{\sigma'} |u_{nk}|^2\\
& \, \leq \,   \sum_{n\in\mathbb{Z}, k\geq 0}
(1+\lambda_k + (\frac{2 \pi n}{\varepsilon})^2)^{\sigma +\sigma'}
|u_{nk}|^2 \\
&\, \leq  \, \|(1-\Delta)^s u\|^2_{L^2}~.
\end{align*}
Finally , we remark that there exists a positive constant $C_6$ 
such that, for $0 < \varepsilon \leq \varepsilon_0$, for
any $u \in X^s$, $\|(1-\Delta)^s u\|_{L^2} \leq 
C_6\|(-\Delta)^s u\|_{L^2}$. Therefore, we deduce from (\ref{patru})
that
\begin{equation}
\label{2embed}
\|u\|_{L^{q,q'}} \, \leq \, \varepsilon^{1/q' -1/2} C_2 
C_0(\varepsilon) C_5 C_6\|u\|_s~,
\end{equation}
which proves the embedding 
$X^s\hookrightarrow L^{q,q'}(Q_\varepsilon)$. If $w \in X^s$ 
satisfies $Mw=0$, then, according to (\ref{2g3}), 
the inequality (\ref{2embed}) becomes
\begin{equation*}
\|u\|_{L^{q,q'}} \, \leq \, \varepsilon^{1/q' -1/2} C_2 
C_1(\varepsilon) C_5 C_6 \|u\|_s \, \leq \, C_2 C_4 C_5 C_6\|u\|_s~,
\end{equation*}
and the estimate (\ref{trei}) is proved.
\end{proof}

In the periodic case, we need an inequality in which the $H^1$
norm of 2-dimensional functions appears. This estimate cannot be
deduced from Lemma \ref{lema1}. We shall show it with the help of the
following commutator estimate.
\begin{lemme}\label{lema3}
Let $\varepsilon_0 > 0$ be fixed.
There exists a positive constant $K_4$ such that, 
for $0 < \varepsilon \leq \varepsilon_0$,
for any functions $f\in \dot{H}_p^1(Q_\varepsilon)$ and 
$g\in \dot{H}_p^{1/2}(Q_\varepsilon)$, where 
  $f$ is independent of $x_3$ and $Mg=0$, the following estimate 
  holds, 
  \begin{equation*}
\nl{[f,(-\Delta)^{1/4}]g}{2} \,\leq \, K_4\ep^{-1/2}\|f\|_1 \|g\|_{1/2},
  \end{equation*}
 where $[f,(-\Delta)^{1/4}]g = f(-\Delta)^{1/4}g - (-\Delta)^{1/4}(fg)$. 
\end{lemme}
\begin{proof}
As in (\ref{Fourier}), we consider the Fourier series of $f$ and $g$:
\begin{equation*}
  f(x) \,=\,  \varepsilon^{-1/2}\sqrt{a_1a_2}
\sum_{m \in {\mathbb Z }^3, m_3= 0} f_m e^{2i\pi ma \cdot x}~, 
\quad g(x) \,=\, \varepsilon^{-1/2}\sqrt{a_1a_2}
\sum_{n \in {\mathbb Z }^3} g_n e^{2i\pi na \cdot x}\,, 
\end{equation*}
where, for $k \in {\mathbb Z }^3$, $ka \equiv (k_1a_1,
k_2a_2,k_3a_3)$ and $a_1=l^{-1}_1$, $a_2=l^{-1}_2$, 
$a_3=\varepsilon^{-1}$. A straightforward computation gives
\begin{equation*}
  [f,(-\Delta)^{1/4}]g \, =\,  \varepsilon^{-1} a_1a_2 \sqrt{2 \pi}
\sum_{m , n \in {\mathbb Z }^3, m_3=0} f_m g_n 
( | n a|^{1/2} - | (n +m)a|^{1/2} )e^{2i\pi (m+n)a \cdot x}~,
\end{equation*}
where $|ka| = (k^2_1a^2_1 + k^2_2a^2_2 + k^2_3a^2_3)^{1/2}$.  Hence,
\begin{equation*}
  \nl{[f,(-\Delta)^{1/4}]g}{2} \, \leq \, \varepsilon^{-1/2} 
\left[ 2 \pi a_1a_2
\sum_{k \in \Z^3} (\sum_{m+n=k, m_3=0}|f_m| |g_n|
\,   |\,|(n +m)a|^{1/2} - | n a|^{1/2} | )^2 \right]^{1/2} .
\end{equation*}
Since $m_3 =0$, there exists a positive constant $C_1$, independent 
of $\varepsilon$, $m$ and $n$, such that 
\begin{equation*}
 |\, | (n +m)a|^{\frac{1}{2}} - | n a|^{\frac{1}{2}}| \, 
 \leq \, C_1  |m^2_1a^2_1 + m^2_2a^2_2|^{\frac{1}{4}}~.
 \end{equation*}
 The previous two inequalities imply that
\begin{equation}
  \label{funu}
  \nl{[f,(-\Delta)^{ 1/4}]g}{2}  \leq 
 C_1 \nl{((-\Delta_2)^{1/4}\check{f})\check{g}}{2},
\end{equation}
where 
\begin{equation*}
\check{f}(x)\, =\, \varepsilon^{-1/2}\sqrt{a_1a_2}
\sum_{m \in {\mathbb Z }^3, m_3=0} |f_m|
e^{2i\pi ma \cdot x}~,\quad
\check{g}(x)\, =\, \varepsilon^{-1/2}\sqrt{a_1a_2}
\sum_{n \in {\mathbb Z }^3} |g_n| e^{2i\pi na \cdot x}~,
\end{equation*}
have the same 
$H^s$-norms as $f$ and $g$ respectively, and where $\check{f}$ is 
independent of $x_3$. H\"older's anisotropic inequality together 
with Lemma \ref{lema2} give
\begin{equation}
  \label{fdoi}
  \begin{split}
  \nl{((-\Delta_2)^{1/4}\check{f})\check{g}}{2} & \, \leq \, 
    \nld{(-\Delta_2)^{1/4}\check{f}}{4}{\Omega} \nl{\check{g}}{4,2} \\
    & \, \leq
  C_2 \nld{(-\Delta_2)^{1/4}\check{f}}{4}{\Omega} \|\check{g}\|_{1/2}
  \, \leq \,  C_2 \nld{(-\Delta_2)^{1/4}\check{f}}{4}{\Omega} 
  \|g\|_{1/2}~.
  \end{split}
\end{equation}
Due to the classical Sobolev embedding $H^{1/2}(\Omega) \hookrightarrow
L^4(\Omega)$, we also have
\begin{equation}
  \label{ftrei}
  \begin{split}
  \nld{(-\Delta_2)^{1/4}\check{f}}{4}{\Omega} & \, \leq \, C_3
  \|(-\Delta_2)^{1/4}\check{f}\|_{H^{1/2}(\Omega)} \, \leq \, C_4
  \|(-\Delta_2)^{1/2}\check{f}\|_{L^2(\Omega)} \\
&\, \leq \,  C_5 \varepsilon^{-1/2} \|\check{f}\|_1
\, \leq \, C_5 \varepsilon^{-1/2} \|f\|_1~.
\end{split}
\end{equation}
We deduce from the relations (\ref{funu}), (\ref{fdoi}) and (\ref{ftrei}) 
that
\begin{equation*}
  \nl{[f,(-\Delta)^{1/4}]g}{2}\, \leq \, C_1 C_2 C_5\varepsilon^{-1/2}
  \|f\|_1 \|g\|_{1/2}.
\end{equation*}
The proof is completed.
\end{proof}

As a consequence, we obtain the following lemma:
\begin{lemme}\label{estimation2d}
Let $\varepsilon_0 > 0$ be fixed.
There exists a positive constant $K_5$ such that, 
for $0 < \varepsilon \leq  \varepsilon_0$, for any vector fields
 $v\in (\dot{H}_p^1(Q_\varepsilon))^3$ and $w\in
  (\dot{H}_p^{3/2}(Q_\varepsilon))^3$, where
$v$ is divergence-free and independent of $x_3$ and  where $Mw=0$,
the following estimate holds
 \begin{equation*}
  \Bigl|\int_{Q_\varepsilon}v(x)\nabla w(x) (-\Delta)^{1/2}w(x)\,dx\Bigl| 
  \,\leq \,  K_5\ep^{-1/2} \|v\|_1 \|w\|_{1/2} \|w\|_{3/2}.
  \end{equation*}
\end{lemme}
\begin{proof}
We can write
\begin{align*}
I & \, \equiv \, \int_{Q_\varepsilon}v(x)\nabla w(x) 
(-\Delta)^{1/2}w(x)\,dx  
\, =\, \int_{Q_\varepsilon}(-\Delta)^{1/4}\bigl(v(x)\nabla w(x)\bigl)
(- \Delta)^{1/4}w(x)\,dx \\
 &\, =\, \int_{Q_\varepsilon}[(-\Delta)^{1/4},v]\nabla w(x)
 (-\Delta)^{1/4}w(x)\,dx+\int_{Q_\varepsilon}v(x)\nabla (-\Delta)^{1/4}w(x)
 (-\Delta)^{1/4}w(x)\,dx~.
\end{align*}
But an integration by parts shows that
\begin{equation*}
  \int_{Q_\varepsilon}v(x)\nabla (-\Delta)^{1/4}w(x) 
  (-\Delta)^{1/4}w(x)\,dx \,
 =\, -\int_{Q_\varepsilon}\operatorname{div}v(x) |(-\Delta)^{1/4}w(x)|^2\,dx=0~.
\end{equation*}
Therefore, we can use Lemma \ref{lema3} to deduce that
\begin{align*}
  |I| &\, =\, \left|\int_{Q_\varepsilon}[(-\Delta)^{1/4},v]\nabla w(x)
  (-\Delta)^{1/4}w(x)\,dx\right| 
  \, \leq \, \nl{[(-\Delta)^{1/4},v]\nabla w}{2}\nl{(-\Delta)^{1/4}w}{2} \\
  &\, \leq \, C_1  \varepsilon^{-1/2}\|v\|_1 \|\nabla w\|_{1/2} \|w\|_{1/2}
 \, \leq \, C_2 \varepsilon^{-1/2}\|v\|_1 \|w\|_{1/2} \|w\|_{3/2}~.
\end{align*}
\end{proof}

\begin{lemme}\label{lemacinq}
Let $\varepsilon_0 > 0$ be fixed.
  There exists a constant $K_6$ independent
  of $\varepsilon$ such that, for $0 < \varepsilon \leq 
  \varepsilon_0$, if $v\in (\dot{H}_p^1(Q_\varepsilon))^3$
 is a divergence-free vector field and
 $v^* \in \dot{H}_p^2(Q_\varepsilon)$ is a function, that are independent of
 $x_3$, then
 \begin{equation}
 \label{com5}
  \Bigl|\int_{Q_\varepsilon}v(x)\nabla v^*(x) \Delta v^*(x)\,dx\Bigl| \, \leq
K_6 \varepsilon^{-1/2} \|v\|_1 \|v^*\|_1 \|v^*\|_2.
 \end{equation}
\end{lemme}
\begin{proof}
Since $v$ is divergence-free, simple integrations by parts give
\begin{equation*}
  \int_{Q_\varepsilon}v(x)\nabla v^*(x) \Delta v^*(x)\,dx
  \, =\, -\sum_{i,j=1}^2 \int_{Q_\varepsilon} \partial_{x_j} v_i(x) 
  \partial_{x_i} v^*(x)  \partial_{x_j} v^*(x) \,dx.
\end{equation*}
We deduce from H\"older's inequality and from a 
two-dimensional Gagliardo-Nirenberg estimate that 
\begin{equation*}
\Bigl|\int_{Q_\varepsilon}v(x)\nabla v^*(x) \Delta v^*(x)\,dx\Bigl| 
\, \leq \,  \varepsilon \|v\|_{H^1(\Omega)} 
\|\nabla v^*\|^2_{L^4(\Omega)}
\, \leq \, C_1\varepsilon \|v\|_{H^1(\Omega)} 
\|v^*\|_{H^1(\Omega)} \|v^*\|_{H^2(\Omega)}~,
\end{equation*}
which implies the lemma.
\end{proof}

\section{A uniqueness result}

The aim of this section is to prove a uniqueness result for weak Leray
solutions. In short, this result says that only the ``purely
3-dimensional'' part of the solution needs to be ``strong'' in order
to obtain uniqueness. In particular, uniqueness of 2D solutions in the
class of 3D weak Leray solutions is obtained. Let us note that, in the case 
of periodic boundary conditions, this
particular fact was already proved by Gallagher \cite{Gallagher97},
while uniqueness of 2D solutions in the class of some ``strong''
solutions was shown by Iftimie \cite{Iftimie9999}.

We start with a remark on the regularity of weak Leray solutions.
\begin{remarque}
\label{rem}
Let $u$ be a weak Leray solution such that $(I-M)u\in
L^\infty(0,T;V_\ep^{1/2})\cap L^2(0,T;V_\ep^{3/2})$. Then $u\in
C^0([0,T];H_{\varepsilon})$ and  $\partial_t u\in 
L^2(0,T;V'_\varepsilon)$.
\end{remarque}
We first show that $u \nabla u$ belongs to $L^2(0,T;V'_\varepsilon)$. Let
$\varphi \in L^2(0,T;V_\varepsilon)$ be a smooth vector in  the $x$ 
variable ($\varphi \in L^2(0,T;V^2_\varepsilon)$ is actually sufficient). 
We have
\begin{equation}
\label{3uphi}
  \int_0^T \int_{Q_\ep} u\nabla u \, \varphi\, dx\,dt\, =\, \int_0^T
 \int_{Q_\ep} (u\nabla Mu \, \varphi + u\nabla (I-M)u \, 
  \varphi)\, dx \,dt~.
\end{equation}
Since $\varphi$ is a smooth vector in the variable $x$, a simple 
integration by parts gives
\begin{equation*}
\int_0^T \int_{Q_\ep} u\nabla Mu \, \varphi\, dx \,dt \, =\, 
 \int_0^T \int_{Q_\ep} u \nabla \varphi \, Mu \, dx \, dt~. 
\end{equation*}
By Lemma \ref{lema1} and Remark \ref{remarque}, we thus 
obtain,
\begin{equation*}
\int_0^T \int_{Q_\ep} u\nabla Mu \, \varphi\, dx \,dt \,
\leq \, C_\ep \left( \int_0^T |u|^2_{1/2} |Mu|^2_{1/2} \,dt \right)^{1/2} 
\|\varphi \|_{L^2(0,T;V_\varepsilon)}~, 
\end{equation*}
and
\begin{align*}
\int_0^T \int_{Q_\ep} u\nabla (I-M)u \varphi\, dx \,dt \,&
\leq \, C_\ep \left(\int_0^T \|u(t)\|^2_{L^2}
  |\nabla (I-M)u(t)|^2_{1/2} \,dt \right)^{1/2} 
  \|\varphi \|_{L^2(0,T;V_\varepsilon)}\, \\
&\leq \, C_\ep \left(\int_0^T \|u(t)\|^2_{L^2} |(I-M)u(t)|^2_{3/2} 
\,dt \right)^{1/2} 
\|\varphi \|_{L^2(0,T;V_\varepsilon)}~.
\end{align*}
By a classical density argument, these estimates are still true for 
any $\varphi \in L^2(0,T;V_\varepsilon)$. We thus conclude that
\begin{multline*}
\left(\int_0^T \|u\nabla u\|^2_{V'_\varepsilon}\,dt \right)^{1/2}
\, \leq \,  C_\ep (\|u\|_{L^4(0,T;V^{1/2}_\varepsilon)} 
 \|Mu\|_{L^4(0,T;V^{1/2}_\varepsilon)}\\
+\|u\|_{L^\infty(0,T;H_\varepsilon)}
\|(I-M)u\|_{L^2(0,T;V^{3/2}_\varepsilon)})~.
\end{multline*}
As $A_\varepsilon u$ and $f$ also belong to 
$L^2(0,T;V'_\varepsilon)$, we infer from the above inequality
that $\partial_t u$ is in the space $L^2(0,T;V'_\varepsilon)$. The 
properties $u \in L^2(0,T;V_\varepsilon)$ and 
$\partial_t u \in L^2(0,T;V'_\varepsilon)$ finally imply that $u$ 
belongs to $C^0([0,T];H_\varepsilon)$.
The proof of the remark is completed.

We can now prove a uniqueness result.

\begin{theoreme}[uniqueness]\label{unic}
  Let $u$ be a weak Leray solution of the Navier-Stokes equations 
  (\ref{NSabs}) such that
$(I-M)u\in L^\infty(0,T;V_\ep^{1/2})\cap L^2(0,T;V_\ep^{3/2})$. 
Then $u$ is unique in the class of the weak Leray solutions.
\end{theoreme}

\begin{proof}
Let $\tilde{u}$ be a weak Leray solution with the same initial data as
$u$. The difference $u-\tilde{u}$ satisfies the following equation in
$V'_\varepsilon$,
\begin{equation}
\label{ec1}
\partial_t(u- \tilde{u})
+\nu A_\varepsilon (u-\tilde{u})+B_\varepsilon (u-\tilde{u}, u)
+B_\varepsilon(\tilde{u},u-\tilde{u}) \, =\, 0~.
\end{equation}
We would like to take the inner product in $L^2(Q_\varepsilon)$ of this equation
with $u-\tilde{u}$ and to integrate in space and time. The result
would be the first line of \iref{fli}. Unfortunately,
this is not possible without some additional justification because the
integral
\begin{equation}
  \label{ec2}
  \int_0^t \int_{Q_\ep} \tilde{u}\nabla(u-\tilde{u}) (u-\tilde{u})\,dx\,d\t,
\end{equation}
which is supposed to vanish, may not converge. Nevertheless, one can 
argue as in \cite{Temam84} and \cite{VonWahl85} (see also
\cite{Gallagher97}). The idea is that, instead of multiplying the
equation of $u-\wt{u}$ by $u-\wt{u}$ which yields regularity problems,
one can multiply the equation of $u$ by $\wt{u}$, the equation of
$\wt{u}$ by $u$ and then subtract the two energy inequalities 
satisfied by $u$ and $\wt{u}$; the result is the same. 
This argument is detailed below.

We saw at the end of the proof of Remark \ref{rem}
that all the terms in the equation of $u$ belong to
$L^2(0,T; V'_\varepsilon)$. So we can multiply the equation of $u$ 
by $\tilde{u} \in L^2(0,T; V_\varepsilon)$ and
integrate in space and time to obtain
\begin{equation}\label{uutil}
 \int_0^t \int_{Q_\ep}( \partial_t u\cdot\wt{u}  +
 \nu \nabla u\cdot\nabla\wt{u} 
 +u\cdot\nabla u \,\wt{u})\, dxd\tau \, =\, 
 \int_0^t \int_{Q_\ep} f\cdot\wt{u}\, dxd\tau~.
\end{equation}

Unfortunately, we cannot directly multiply the equation of
$\tilde{u}$ by $u$ and then integrate in space and time, because 
 $\partial_t \wt{u}$ and $u$ are only in $L^{4/3}(0,T;V'_\varepsilon)$
and $L^{2}(0,T;V_\varepsilon)$ respectively. As $u\in
C^0([0,T];H_\varepsilon)\cap L^2(0,T;V_\varepsilon) 
\cap H^1(0,T;V'_\varepsilon)$, by a standard smoothing procedure, 
we can find a sequence of smooth divergence free 
vector fields $u_n \in V_\varepsilon$, such
that $u_n$ converges strongly to $u$ in $C^0([0,T];H_\varepsilon)
\cap L^2(0,T;V_\varepsilon) \cap L^4(0,T;V^{1/2}_\varepsilon)$,
$\partial_t u_n$ converges strongly to $\partial_t u$ in
$L^2(0,T;V'_\varepsilon)$ and $(I-M)u_n$ converges strongly to 
$(I-M)u$ in $L^2(0,T;V^{3/2}_\varepsilon)$. Multiplying the 
equation of $\wt{u}$ by $u_n$  and integrating by parts yield
\begin{equation}\label{lim1}
 \int_0^t \int_{Q_\ep} (\partial_t \wt{u}\cdot u_n  +
 \nu \nabla \wt{u} \cdot \nabla u_n 
 -\wt{u}\cdot\nabla u_n \,\wt{u} )\,  dxd\tau =
\int_0^t \int_{Q_\ep} f \cdot  u_n \, dxd\tau~.
\end{equation}
We now pass to the limit in $n$ in the above equation. With the 
regularities and convergences at hand, it is easily seen that
\begin{equation}
\label{lim2}
 \int_0^t \int_{Q_\ep} \nabla \wt{u}\nabla u_n \, dxd\tau
 \rightarrow \int_0^t \int_{Q_\ep} \nabla \wt{u}\nabla u \,
 dxd\tau \quad \text{and} \quad
 \int_0^t \int_{Q_\ep} f u_n \, dxd\tau \rightarrow 
 \int_0^t \int_{Q_\ep} f u \, dxd\tau~.
\end{equation}

On the other hand, by Lemma \ref{lema1}, we have
\begin{align*}
\int_0^t \int_{Q_\ep}\wt{u}\cdot\nabla(u-u_n) \wt{u}\, dxd\tau  \, 
 &= \, \int_0^t \int_{Q_\ep}(\wt{u}\cdot\nabla M(u-u_n) \, \wt{u}
 +\wt{u}\cdot\nabla (I-M)(u-u_n) \, \wt{u})\,dxd\tau \, \\
 &\, \leq \, C_\varepsilon \int_0^t |\wt{u}|^2_{1/2} 
 (|M(u-u_n)|_{1}+|(I-M)(u-u_n)|_{3/2}) \, d\tau \\
 &\, \leq C_\varepsilon \|\wt{u}\|^2_{L^4(0,T;V^{1/2}_\varepsilon)} 
 \bigl(\|Mu-Mu_n\|_{L^2(0,T;V_\varepsilon)} \\
 & \phantom{\, \leq C\|\wt{u}\|^2_{L^4(0,T;V^{1/2}_\varepsilon)}\, }+ \|(I-M)u-(I-M)u_n\|_{L^2(0,T;V^{3/2}_\varepsilon)}\bigr).
\end{align*}
We deduce that
\begin{equation}\label{lim3}
 \int_0^t \int_{Q_\ep}\wt{u}\cdot\nabla u_n \,\wt{u} \, 
 dxd\tau \rightarrow
 \int_0^t \int_{Q_\ep}\wt{u}\cdot\nabla u \,\wt{u} \, dxd\tau~.
\end{equation}
Finally, we integrate by parts to obtain that
\begin{equation*}
  \int_0^t \int_{Q_\ep} \partial_t \wt{u}\cdot u_n \, dxd\tau \,
  =\, -\int_0^t \int_{Q_\ep}  \wt{u}\cdot \partial_t u_n \, dxd\tau
  +\int_{Q_\ep} (\wt{u}(t)\cdot u_n(t) - \wt{u}(0)\cdot u_n(0)) dx~.
\end{equation*}
As $\partial_t u_n$ and $u_n$ converge 
to $\partial_t u$ and $u$ in $L^2(0,T;V'_\varepsilon)$ and 
$C([0,T];H_\varepsilon)$ respectively, we infer from the above 
equality that,
\begin{equation}
\label{lim4}
  \int_0^t \int_{Q_\ep} \partial_t \wt{u}\cdot u_n\, dxd\tau \rightarrow
  -\int_0^t \int_{Q_\ep}  \wt{u}\cdot \partial_t u \, dxd\tau
  +\int_{Q_\ep} (\wt{u}(t)\cdot u(t) -\wt{u}(0)\cdot u(0))\, dx~.
\end{equation}

Putting together the properties \iref{lim1}, \iref{lim2}, \iref{lim3} and
\iref{lim4}  finally yields
\begin{multline}
  \label{utilu}
\int_0^t \int_{Q_\ep} (- \wt{u}\cdot \partial_t u +
  \nu \nabla \wt{u}\cdot\nabla u
 -\wt{u}\cdot\nabla u \,  \wt{u}\,)dxd\tau \, \\
  =\, -\int_{Q_\ep} (\wt{u}(t)\cdot u(t) - \wt{u}(0)\cdot u(0)) dx
  +\int_0^t \int_{Q_\ep} f\cdot u \, dxd\tau~.
\end{multline}
Since $u$ and $\wt{u}$ are weak Leray solutions, the two following energy
inequalities hold:
\begin{gather*}
 \frac{1}{2} \|u(t)\|^2_{L^2}+ \nu \int_0^t \|\nabla u\|^2_{L^2}d\tau  \,
 \leq \,\frac{1}{2}\|u_0\|^2_{L^2}+  \int_0^t \int_{Q_\ep} f\cdot u\, 
 dx d\tau~,\\
 \frac{1}{2} \|\wt{u}(t)\|^2_{L^2}+ \nu \int_0^t \|\nabla 
 \wt{u}\|^2_{L^2}d\tau  \,
 \leq \,\frac{1}{2}\|u_0\|^2_{L^2}+  \int_0^t \int_{Q_\ep} f\cdot 
 \wt{u} \, dx d\tau~.
\end{gather*}
We now add both energy inequalities and subtract relations
\iref{uutil} and \iref{utilu} to obtain
\begin{equation*}
  \frac{1}{2}\nl{u(t)- \tilde{u}(t)}{2}^2+ \nu\int_0^t 
  \|u-\tilde{u}\|_1^2 d\tau \,
  \leq \, \int_0^t \int_{Q_\ep}(-\wt{u}\cdot\nabla u \, \wt{u}
  + u\cdot\nabla u \,\wt{u})dx d\tau~.
\end{equation*}
Arguing as in Remark \ref{rem}, one shows that the integral
$\int_0^t \int_{Q_\ep} (-u\cdot\nabla u\, u 
+ \tilde{u}\cdot\nabla u\, u ) dxd\tau=0$ is absolutely convergent 
and vanishes. Thus we deduce from the previous inequality that
\begin{equation}
\label{fli}
\nl{u(t)- \tilde{u}(t)}{2}^2+ 2\nu\int_0^t \|u-\tilde{u}\|_1^2\,d\t 
\, \leq \, 
 -2\int_0^t \int_{Q_\ep} (u-\tilde{u})\nabla u (u-\tilde{u})dx d\tau~.
 \end{equation}
Writing $\nabla u = \nabla ((I-M)u + Mu)$ and applying Lemma \ref{lema1}, 
we get, for any $0<s<1$,
\begin{multline*}
  \nl{u(t)- \tilde{u}(t)}{2}^2 +2\nu \int_0^t \|u-\tilde{u}\|_1^2\,d\t
  \leq C_1\int_0^t\nl{u-\tilde{u}}{2} \|u-\tilde{u}\|_1
\|(I-M)u\|_{3/2}\,d\t \\
  + C_2\int_0^t \|Mu\|_1 \|u-\tilde{u}\|_s \|u-\tilde{u}\|_{1-s}\,d\t~.
  \end{multline*}
 Since the interpolation inequality $\|u-\tilde{u}\|_{\tilde{s}} \leq C_3
 \|u-\tilde{u}\|_1^{\tilde{s}} \nl{u-\tilde{u}}{2}^{1-\tilde{s}}$ holds,
for any
 $\tilde{s} \in [0,1]$, we infer from the above inequality that
\begin{multline*}
  \nl{u(t)- \tilde{u}(t)}{2}^2 + 2\nu \int_0^t \|u-\tilde{u}\|_1^2\,d\t\,
 \leq \, 2\nu \int_0^t \|u-\tilde{u}\|_1^2\,d\t \\
  + C_4\int_0^t \nl{u-\tilde{u}}{2}^2(
\|(I-M)u\|_{3/2}^2 +
 \|Mu\|_1^2)\,d\t~,
\end{multline*}
that is
\begin{equation*}
  \nl{u(t)- \tilde{u}(t)}{2}^2 \,
 \leq \, C_4\int_0^t \nl{u-\tilde{u}}{2}^2(
\|(I-M)u\|_{3/2}^2 +
 \|Mu\|_1^2)\,d\t~.
\end{equation*}
And the result follows from Gronwall's inequality.
\end{proof}

\section{The case of mixed boundary conditions}

In this section, we shall prove the theorems \ref{theoreme1} and
\ref{theoreme1bis}.

\begin{proof}[Proof of Theorem \ref{theoreme1}]
The proof is based on a Galerkin approximation, using the first $m$
eigenvectors $\psi_1$, $\psi_2$, ....,$\psi_m$
of the Stokes operator $A_\varepsilon$. Since $M$
and $A_\varepsilon$ commute, we can choose these eigenvectors
 $\psi_j$ so that, either $\psi_j \in MV^2_\varepsilon$ or $\psi_j \in
 (I-M)V^2_\varepsilon$.
Let $\mathcal{P}_m: H_\varepsilon \rightarrow H_\varepsilon$ denote the
projector onto the space $\mathcal{V}_m$ generated
by the first $m$ eigenfunctions. We remark that $\mathcal{P}_m M
=M \mathcal{P}_m$. The above properties imply that, for every
$s \in [0,2]$ and for every $u \in V^s_\varepsilon$,
\begin{equation}
\label{4Pm}
|\mathcal{P}_m (I-M)u|_s \, \leq \, |(I-M)u|_s~, \quad
|\mathcal{P}_m Mu|_s\, \leq \, |Mu|_s~.
\end{equation}

We know (see \cite{ConstantinFoias}, Chapter 8, for example or 
\cite{Temam84}), that, for
every $m \in \mathbb{N}$, there exists a global solution
$u_m \in C^1([0,+\infty); V^2_\varepsilon \cap \mathcal{V}_m)$ of 
the equations (\ref{NSabs}) or
also of (\ref{NSabsv}) and (\ref{NSabsw}), where $B_\varepsilon$ is
replaced by $\mathcal{P}_m
B_\varepsilon$ and $P_\varepsilon f$ by $\mathcal{P}_m P_\varepsilon f$ and 
where the initial
condition is $u_m(0)=\mathcal{P}_m (I-M)u_0 + \mathcal{P}_m Mu_0 \equiv
w_{0m} + v_{0m}$.
Moreover, for every $\tau >0$, $u_m$ 
and $\partial_t u_m$ are uniformly bounded with respect
to $m$ in the spaces $L^{\infty}(0, +\infty;H_\varepsilon) \cap 
L^2(0, \tau;V_\varepsilon)$ and $L^{4/3}(0, \tau;V'_\varepsilon)$ 
respectively. We want to show that
this solution $u_m \equiv  w_m + v_m  = (I-M)u_m + Mu_m$ satisfies 
the additional estimates and properties
given in Theorem \ref{theoreme1}, which will be preserved, when $m$ 
goes to $+\infty$. In order to simplify the notation, we shall drop the 
subscript $m$ in all the a priori estimates below, when there is no confusion.
Taking the inner product of the modified equation (\ref{NSabsw}) with
$A^{1/2}w$ gives, for $t \geq 0$,
\begin{multline}
\label{ff0}
  \frac{1}{2} \partial_t |w(t)|_{1/2}^2 +\nu|w(t)|_{3/2}^2
+\int_{Q_\varepsilon}
  (w\nabla w \, (I-M)A^{1/2}w)(t,x) dx +\int_{Q_\varepsilon} (v\nabla
  w \, A^{1/2}w)(t,x) dx\\
  +\int_{Q_\varepsilon} (w\nabla v \, A^{1/2}w)(t,x) dx \,= \,
\int_{Q_\varepsilon}
  ((I-M)Pf \,  A^{1/2}w)(t,x) dx.
\end{multline}
Since $I-M$ commutes with $A^{1/4}$, we get by 
Lemma \ref{lema1}, for $t \geq 0$,
\begin{align*}
  \left|\int_{Q_\varepsilon} (w\nabla w (I-M)A^{1/2}w)(t,x) dx\right|
  &\, \leq \, C|w(t)|_1 \|\nabla w(t)\|_{L^2} |(I-M)A^{1/2}w(t)|_{1/2}\\
  & \, \leq \, C|w(t)|_1^2|w(t)|_{3/2}~.
\end{align*}
A simple interpolation inequality now yields, for $t \geq 0$,
\begin{equation}
  \label{ff1}
 \begin{split}
  \left|\int_{Q_\varepsilon} (w\nabla w (I-M)A^{1/2}w)(t,x) dx\right|
& \,  \leq \, C|w(t)|_{1/2} |w(t)|^2_{3/2} \\
&\,  \leq \, C\nu^{-1}|w(t)|_{1/2}^2 |w(t)|_{3/2}^2 +
\frac{\nu}{8}|w(t)|_{3/2}^2~.
\end{split}
\end{equation}
The inequality (\ref{doi}) of Lemma \ref{lema1} implies, for $s\in [1/2,1)$,
\begin{equation*}
\begin{split}
  \left|\int_{Q_\varepsilon} (v\nabla w \, A^{1/2}w)(t,x) dx\right|
&\,  \leq \, C_s\ep^{-1/2} |v(t)|_s \nl{\nabla w(t)}{2}
|A^{1/2}w(t)|_{1-s} \\
& \, \leq \, C_s\ep^{-1/2} |v(t)|_s |w(t)|_{2-s} |w(t)|_{1}~,
 \end{split}
\end{equation*}
where $C_s$ denotes a positive constant depending only $s$.
We find again by interpolation that, for $t \geq 0$,
\begin{equation}
  \label{ff2}
  \begin{split}
  \left|\int_{Q_\varepsilon} (v\nabla w \, A^{1/2}w)(t,x) dx\right|
  &\, \leq \, C_s \ep^{-1/2} |v(t)|_s |w(t)|_{1/2}^s |w(t)|_{3/2}^{2-s}\\
  &\, \leq C_s\, \nu^{1-2/s}\ep^{-1/s} |v(t)|_s^{2/s} |w(t)|_{1/2}^2
  +\frac{\nu}{8}|w(t)|_{3/2}^2~.
 \end{split}
\end{equation}
Due to the estimate (\ref{doi}) of Lemma \ref{lema1}, we also have,
for $t \geq 0$,
\begin{equation}
  \label{ff3}
  \begin{split}
  \left|\int_{Q_\varepsilon} (w\nabla v \, A^{1/2}w)(t,x) dx \right|
  &\, \leq \, C\ep^{-1/2}|w(t)|_{1/2} \nl{\nabla v(t)}{2} |A^{1/2}w(t)|_{1/2}\\
  &\, \leq \, C\ep^{-1/2} |v(t)|_1 |w(t)|_{1/2} |w(t)|_{3/2}\\
  &\, \leq C\, \ep^{-1}\nu^{-1} |v(t)|_1^2 |w(t)|_{1/2}^2
  +\frac{\nu}{8}|w(t)|_{3/2}^2~.
  \end{split}
\end{equation}
Finally, we obtain, due to (\ref{4Pm}) and the Poincar\'{e} inequality
(\ref{Poincarewbis}) that, for $t \geq 0$,
\begin{equation}
  \label{ff4}
  \begin{split}
  \left| \int_{Q_\varepsilon} ( (I-M)Pf \, A^{1/2}w)(t,x) dx \right|
   &\, \leq \, C\nl{\mathcal{P}_m (I-M)P  f (t)}{2} |w(t)|_1\\
  &\, \leq \, C\ep^{1/2} \nl{(I-M)P  f(t) }{2} |w(t)|_{3/2}\\
  &\, \leq \, C\nu^{-1}\ep \nl{(I-M)P  f(t) }{2}^2 +
\frac{\nu}{8}|w(t)|_{3/2}^2~.
  \end{split}
\end{equation}
We now fix the real number $s \in [1/2,1)$. In particular, we assume
that $s$ satisfies the following conditions
\begin{equation}
\label{4conds}
1\, > \, s\, > \, \sup ( \beta, 2\gamma, \frac{1}{2}, \frac{\alpha}{\alpha +
1},
\frac{\alpha}{\alpha +1 -\beta}, \frac{1}{1 +
\delta})~.
\end{equation}
The inequalities  (\ref{ff1}), (\ref{ff2}), (\ref{ff3}) and
(\ref{ff4}), together with (\ref{ff0}), imply, for $t \geq 0$,
\begin{multline}
\label{4resume1}
  \partial_t |w(t)|_{1/2}^2 +\frac{\nu}{2}|w(t)|_{3/2}^2 \,\leq\,
 C_1|w(t)|_{1/2}^2\bigl(\nu^{1-2/s}\ep^{-1/s} |v(t)|_s^{2/s}
  +\nu^{-1}\ep^{-1} |v(t)|_1^2
  +\nu^{-1}|w(t)|_{3/2}^2\bigl)\\
  +C_1\nu^{-1}\ep\nl{(I-M)P  f (t)}{2}^2~.
\end{multline}
Due to the property (\ref{4Pm}) and
to the hypothesis (\ref{10}) on the initial data,
when $K_{\ast}$ is small enough, there exists a
positive time $T$ such that, for $t \in [0,T)$,
\begin{equation}
  \label{4Hwv0}
|w(t)|^2_{1/2}\, < \frac{\nu^2}{4C^2_1}~,
\end{equation}
and, that, if $T < \infty$,
\begin{equation}
  \label{4Hwv0egal}
|w(T)|^2_{1/2} \,=\, \frac{\nu^2}{4C^2_1}~.
\end{equation}
We shall show by contradiction that $T=+\infty$.
We derive from (\ref{4resume1}), (\ref{4Hwv0}),
and (\ref{4Hwv0egal}), that, for $t \in [0,T]$,
\begin{multline}
  \label{cincistar}
  \partial_t |w(t)|_{1/2}^2 +\frac{\nu}{4}|w(t)|_{3/2}^2 \, \leq\,
 C_1 |w(t)|_{1/2}^2\bigl(\nu^{1-2/s}\ep^{-1/s} |v(t)|_s^{2/s}
  +\nu^{-1}\ep^{-1} |v(t)|_1^2 \bigl) \\
  +C_1\nu^{-1}\ep\nl{(I-M)P  f (t)}{2}^2~.
\end{multline}
which in turn implies
\begin{multline}
  \label{alta}
  \partial_t |w(t)|_{1/2}^2 +\frac{\nu\ep^{-2} K_0^{-2}}{4}|w(t)|_{1/2}^2 \,
  \leq\,
  C_1|w(t)|_{1/2}^2\bigl(\nu^{1-2/s}\ep^{-1/s} |v(t)|_s^{2/s}
  +\nu^{-1}\ep^{-1} |v(t)|_1^2 \bigl) \\
  +C_1\nu^{-1}\ep\nl{(I-M)P  f(t) }{2}^2~.
\end{multline}
Set
\begin{equation*}
\begin{split}
  h^\ast(t)&\, =\, C_1\nu^{1-2/s}\ep^{-1/s} \int_0^t |v(\t)|_s^{2/s}d\t+
   C_1\nu^{-1}\ep^{-1} \int_0^t |v(\t)|_1^2d\t
   -\nu t K^{-2}_0\varepsilon^{-2}/4 \\
 &\, \equiv \, h(t)  -\nu t K^{-2}_0\varepsilon^{-2}/8 ~.
 \end{split}
\end{equation*}
An application of Gronwall's lemma in (\ref{alta}) gives, for $0 \leq
t \leq T$,
\begin{multline}
\label{deca}
|w(t)|_{1/2}^2 \, \leq \,  \exp (h^\ast(t)) \, |w_0|_{1/2}^2
  + C_1 \varepsilon \nu^{-1}
  \int_0^t\|(I-M)P f (\tau)\|^{2}_{L^2} \exp (h^\ast(t)-h^\ast(\tau)) d\tau \\
 \, \leq \,  \exp (h^\ast(t)) \, |w_0|_{1/2}^2 + 8\nu^{-2} K^2_{0} C_1
 \varepsilon^3
 ( \sup_t  \|(I-M)P f (t)\|^{2}_{L^2})(\sup_{0 \leq \tau \leq t}
  \exp (h(t)-h(\tau)))~.
 \end{multline}
 The estimate of $h(t)-h(\tau)$ is simple and comes
from the usual $L^2$-energy estimates on the velocity $u$. If we
take the inner product of the modified equation (\ref{NSabs}) with
$u$, we obtain, for $t \geq 0$,
\begin{equation}
\label{4dL2u}
\begin{split}
  \partial_t \nl{u(t)}{2}^2 +2\nu|u(t)|_1^2 &\, =\, 2
  \int_{Q_\varepsilon} (P f \cdot u)(t,x)\,dx \\
 & \, \leq \, 2 (\| (I-M) Pf(t)\|_{L^2} \|w(t)\|_{L^2} +
  \|MPf(t)\|_{L^2} \|v(t)\|_{L^2})\\
&\, \leq \,  \nu^{-1} (K_0 \varepsilon \| (I-M) Pf(t)\|_{L^2} + \mu_0
\|MPf(t)\|_{L^2})^2 + \nu |u(t)|^2_1 ~.
\end{split}
\end{equation}
It follows that, for $t \geq 0$,
\begin{equation}
  \label{fff1}
 \partial_t \nl{u(t)}{2}^2 +\nu \mu^{-2}_0 \|u(t)\|^2_{L^2} \, \leq \,
\partial_t \nl{u(t)}{2}^2 +\nu |u(t)|^2_1
 \, \leq \, \nu^{-1} B~,
\end{equation}
 where
 $B = 2( \mu^2_0 \sup_t \|MPf(t)\|^2_{L^2} + \varepsilon^2 K^2_0
 \sup_t \|(I-M)Pf(t)\|^2_{L^2})$. Gronwall's lemma implies, for $t 
 \geq 0$,
\begin{equation}
\label{4u0L2}
  \nl{u(t)}{2}^2 \, \leq \,
  \nl{u_0}{2}^2 + \nu^{-2}\mu_0^2 B(1- e^{-\nu \mu^{-2}_0 t})~.
\end{equation}
Integrating (\ref{fff1}) from $t_0$ to $t_1$, where $0 \leq t_0 \leq
t_1$, one finds
\begin{equation}
\label{unustar}
\begin{split}
  \nl{u(t_1)}{2}^2 +\nu \int_{t_0}^{t_1} |u(\t)|_1^2 d\t
& \, \leq \, \nl{u(t_0)}{2}^2 +\nu^{-1} (t_1-t_0) B  \\
& \, \leq \,
  \nl{u_0}{2}^2 +\nu^{-1}B (\nu^{-1}\mu_0^2
  (1- e^{-\nu \mu^{-2}_0 t_0}) + (t_1 - t_0))~.
\end{split}
\end{equation}
By interpolation, we can write,
\begin{equation}
  \label{ff5}
 \int_{t_0}^{t_1}  |v(\tau)|_s^{2/s}d\tau \,
\leq \, C_2\sup_{t_0 \leq  \tau \leq t_1} \nl{v(\tau)}{2}^{2(1-s)/s}
  \int_{t_0}^{t_1} |v(\tau)|_1^2 d\tau~.
\end{equation}
Since $M$ is an orthogonal projection in $H_\varepsilon$ and
$V_\varepsilon$, we infer from (\ref{4u0L2}), (\ref{unustar}) and
(\ref{ff5}) that, for $0 \leq t_0 \leq t_1$,
\begin{multline}
\label{rica}
  h(t_1)- h(t_0) \, \leq \, C_3\Bigl(\varepsilon^{-1}
  + \varepsilon^{-1/s} \bigl(\nl{u_0}{2}^2 + t^\ast_1 B
  \bigl)^{(1-s)/s}\Bigl)\\
  \times\Bigl(\nl{u_0}{2}^2 + (t_1-t_0 +t^\ast_0) B \Bigl)
  -\nu (t_1-t_0)\varepsilon^{-2}K^{-2}_0/8~,
\end{multline}
where $t^\ast= \min(1,t)$. To simplify, we set $\|v_0\|_{L^2}= K_\ast
\varepsilon^{1/2} g_0(\varepsilon)$ and $\sup_t \|MP_\varepsilon 
f(t)\|_{L^2} =K_\ast \varepsilon^{1/2}g_1(\varepsilon)$, where $g_0$ 
and $g_1$ are positive functions of $\varepsilon$.
The hypotheses (\ref{10}) give the following bounds,
\begin{equation}
\label{4bornhyp}
\begin{split}
  \nl{u_0}{2}^2 &\, =\, \nl{v_0}{2}^2+\nl{w_0}{2}^2\,
  \leq \, K^2_\ast \varepsilon(K^2_0 + g^2_0) \, \leq \,
  K^2_\ast \varepsilon(K^2_0 + \varepsilon^{-2\alpha})~,\\
B & \, \leq \, C_4 K^2_\ast \varepsilon ( |\ln \varepsilon|^{2\gamma}
+ g^2_1) \, \leq \, C_4 K^2_\ast \varepsilon
( |\ln \varepsilon|^{2\gamma} + \varepsilon^{-2\beta})~.
\end{split}
\end{equation}
We now deduce from \iref{rica} and (\ref{4bornhyp}) that,
for $0 \leq t_0 \leq t_1 $,
\begin{equation}
\label{4ht1t0}
\begin{split}
 h(t_1)- h(t_0) & \,  \leq \,  C_5  K^2_\ast (1 +  K^2_\ast)
 \left(1+ g^{2/s}_0 + (t^\ast_1 + t^\ast_0)(g^{2/s}_1 +|\ln
 \varepsilon|^{2\gamma/s}) \right) \\
  - ( & t_1-t_0) \Bigl( \frac{\nu K^{-2}_0}{8} \varepsilon^{-2} -
 C_5  K^2_\ast (1 +  K^2_\ast)(g^{2/s}_1 +|\ln
 \varepsilon|^{2\gamma/s} + g^{\frac{2(1-s)}{s}}_0(g^2_1 +|\ln
 \varepsilon|^{2\gamma}) \Bigr)~.
 \end{split}
 \end{equation}
 Due to the choice (\ref{4conds}) of $s$ and the hypotheses
 (\ref{10}) and (\ref{10bis}), we infer from (\ref{4ht1t0}) that, 
 when $K_\ast$ is small enough, we have, for $0 \leq t_0 \leq t_1$,
 \begin{equation}
 \label{4ht1t0fin}
h(t_1)- h(t_0) \,  \leq \,  C_6  K^2_\ast (1+ g^{2/s}_0 + g^{2/s}_1
+|\ln  \varepsilon|^{2\gamma/s})~.
\end{equation}
Likewise, we derive from (\ref{4ht1t0}) that, when
$K_\ast$ is small enough, we have, for $t \geq 0$,
 \begin{equation}
 \label{4ht}
h(t) \,  \leq \,  C_7  K^2_\ast (1+ g^{2/s}_0 )~.
\end{equation}
Finally, we deduce from (\ref{4ht1t0fin}), (\ref{4ht}), (\ref{deca}),
(\ref{4conds}) and the hypotheses (\ref{10}), (\ref{10bis}),
where $K_\ast$ is small enough, that, for
$0 \leq t \leq T$,
\begin{equation}
\label{4wfin}
\begin{split}
 |w(t)|_{1/2}^2\, & \leq \, C_9\exp (C_8 K^2_\ast g^{2(1+\delta)}_0) 
 \Bigl( \exp (-\frac{\nu t K^{-2}_0}{8} \varepsilon^{-2}) 
 |w_0|_{1/2}^2 \\
& \phantom{\leq \, C_9\exp (C_8 K^2_\ast g^{2(1+\delta)}_0)}
+ \varepsilon^3 \sup_t \|(I-M)Pf(t)\|^2_{L^2}
\exp (C_8 K^2_\ast (g^{2(1+\delta)}_1 +
|\ln  \varepsilon|^{\frac{2\gamma}{s}}) \Bigr) \\
& \leq \,  C_{10} K_{\ast} \, \leq \, \frac{\nu^2}{8C^2_1}~,
\end{split}
\end{equation}
which contradicts the property (\ref{4Hwv0egal}), if $T < +\infty$. 
It follows that $T =+\infty$. Remark that the estimate 
(\ref{4u0L2}) implies, for $t \geq 0$,
\begin{equation}
\label{4vfin}
  \nl{v(t)}{2}^2 \, \leq \,
  \nl{u_0}{2}^2 + \nu^{-2}\mu_0^2 B \, \leq \, 
  \nl{u_0}{2}^2 + C_{11}\varepsilon^{-1}~.
\end{equation}

We have just proved that, under the hypotheses
(\ref{10}) and (\ref{10bis}), for any $m \in \mathbb{N}$, the
solution $u_m \in C^1([0, +\infty); \mathcal{V}_m)$ of the modified
Navier-Stokes equations
(\ref{NSabs}) with initial data $u_m(0)= \mathcal{P}_m u_{0m}$
satisfies
\begin{equation}
\label{4inegam1}
\sup_{t \geq 0} |w_m(t)|^2_{1/2} \, <\, C_{12}~,
\end{equation}
where $C_{12}$ is a positive constant independent of
$\varepsilon$ and $m$. Integrating the inequality (\ref{cincistar})
from $0$ to $t$ and using the estimates (\ref{4inegam1}),
(\ref{unustar}) and (\ref{ff5}),
one also shows that, for any $t \in [0,+ \infty)$,
\begin{equation}
\label{4inegam2}
\int_{0}^{t} |w_m(s)|^2_{3/2} d{s} \,
\leq \, C_{13}(\varepsilon) t~,
\end{equation}
where $C_{13}(\varepsilon)$ is a positive constant independent of
$m$, but depending on $\varepsilon$.

We remark that $v_{0m}$ and $w_{0m}$ converge to $Mu_0$ and 
$(I-M)u_0$ in $H_{\varepsilon}$ and $V^{1/2}_\varepsilon$ respectively.
Now, a classical argument (see \cite{ConstantinFoias},
Chapter 8 or \cite{Temam84}) shows that $u =\lim_{m \rightarrow
+\infty} u_m$ belongs to the space
$L^{\infty}(0,\infty; H_\varepsilon)
\cap L^2_{loc}([0,\infty); V^1_\varepsilon)$, is a weak Leray
solution of the equations (\ref{NSabs}) with initial data $u(0)=u_0$
and that, due to the properties (\ref{4inegam1}) and 
(\ref{4inegam2}),
$(I-M)u$ belongs to $L^\infty(0,\infty;V^{1/2}_\varepsilon) \cap
L^2_{loc}([0,\infty);V^{3/2}_\varepsilon)$. The
uniqueness of the solution $u$ follows from Theorem \ref{unic}.
Remark \ref{rem} implies that $u$ belongs to the space
$C^0([0,+\infty); H_\varepsilon) \cap 
L^2_{loc}([0,+\infty); V'_{\varepsilon})$.
\end{proof}

\begin{remarque}\label{Remark41}
In the (PP) case, we can apply Lemma \ref{estimation2d} in order to
estimate the term $|\int_{Q_\varepsilon} (v\nabla w A^{1/2}w)(t,x)
dx|$, which gives
\begin{equation*}
|\int_{Q_\varepsilon} (v\nabla w A^{1/2}w)(t,x) dx| \, \leq \,
C \nu^{-1} \varepsilon^{-1} |v(t)|_1^2 |w(t)|_{1/2}^2
  +\frac{\nu}{8}|w(t)|_{3/2}^2~.
 \end{equation*}
In this case, $h(t)=  C_1\nu^{-1}\ep^{-1} \int_0^t |v(\t)|_1^2d\t
  -\nu t K^{-2}_0\varepsilon^{-2}/8$ and the estimate of
  $h(t_1)-h(t_0)$ becomes
  \begin{equation}
 \label{4ht1t0per}
 \begin{split}
 h(t_1)- h(t_0) \,  \leq \, & C_5  K^2_\ast (1 +  K^2_\ast)
 \left(1+ g^2_0 + (t^\ast_1 + t^\ast_0)(g^{2}_1 +|\ln
 \varepsilon|^{2\gamma}) \right) \\
 & - (t_1-t_0) \Bigl( \frac{\nu K^{-2}_0}{8} \varepsilon^{-2} -
 C_5  K^2_\ast (1 +  K^2_\ast)(g^{2}_1 +|\ln
 \varepsilon|^{2\gamma}) \Bigr)~.
 \end{split}
 \end{equation}
Hence, we can choose $\beta =1$, $\gamma =1/2$ in the hypothesis
(\ref{10}). Moreover, the limitation on $Mu_0$ disappears.
The condition (\ref{10bis}) then writes as (\ref{10bisbis}).
\end{remarque}

We now prove Theorem \ref{theoreme1bis}.
\begin{proof}[Proof of Theorem \ref{theoreme1bis}]
The proof follows the same lines as the proof of Theorem
\ref{theoreme1}. So we shall only indicate the main changes in the
estimate of $|w(t)|_{1/2}$, for $0 \leq t \leq T$. Let $s \in
[1/2,1)$ be fixed so that $s > \frac{1}{1 + \delta}$. Arguing as
in (\ref{fff1}) and (\ref{4u0L2}), we deduce from (\ref{4dL2u})
that, for $t \geq 0$,
\begin{equation}
\label{4uL2H1bis}
\|u(t)\|^2_{L^2} + \nu \int_{0}^{t}|u(\tau)|^2_1 d\tau \, \leq \, D~,
\end{equation}
where $D=\|u_0\|^2_{L^2} + C_{14} \int_{0}^{t}
(\|MPf(\tau)\|^2_{L^2} + \varepsilon^2 \|(I-M)Pf(\tau)\|^2_{L^2} )
d\tau$. The hypothesis (\ref{10ter}) imply that
\begin{equation}
\label{4bornD}
D \, \leq \, C_{15}  \varepsilon (\tilde{K} + \varepsilon^{-1} 
\|Mu_0\|^2_{L^2} + \varepsilon^{-1} \int_{0}^{t} \|MPf(\tau)\|^2_{L^2}
d\tau)~.
\end{equation}
The application of Gronwall's lemma to (\ref{alta}) and the estimate
(\ref{4uL2H1bis}) give, for $0 \leq t \leq T$,
\begin{equation}
\label{decabis}
|w(t)|^2_{1/2} \, \leq \, (\exp \, C_{16} ( \varepsilon^{-1} D + 
(\varepsilon^{-1} D)^{1/s})) (|w_0|^2_{1/2}+
\varepsilon C_1 \nu^{-1} \int_{0}^{+\infty} \|(I-M)Pf(\tau)\|^2_{L^2}
d \tau)~,
\end{equation}
which implies, due to (\ref{4bornD}) and the hypothesis
(\ref{10ter}), where $\tilde{K}$ is small enough, 
that, for $0 \leq t \leq T$,
\begin{equation}
|w(t)|^2_{1/2} \, < \, \frac{\nu^2}{8 C^2_1}~.
\end{equation}
Now we finish the proof by arguing like in the proof of
Theorem \ref{theoreme1}.
\end{proof}

\section{The case of periodic boundary conditions}

In the periodic case, we obtain better results than those described in
Theorem \ref{theoreme1} because we can
use the conservation of enstrophy property, which is valid for
two-dimensional periodic vector fields.  We recall that, for this reason, 
we split the vector field $v \equiv Mu$ into two parts
\begin{equation*}
Mu \, = \, M\tilde{u} + M(u_3)\, \equiv \, (Mu_1,Mu_2,0) + (0,0, Mu_3)~,
\end{equation*}
and set $\tilde{v} = M\tilde{u}$.
Likewise, we shall split the forcing term as follows
\begin{equation*}
M(Pf) \, = \, M\widetilde{Pf} + M((Pf)_3)\, \equiv  \,
(M(Pf)_1,M(Pf)_2,0) + (0,0,M(Pf)_3)~.
\end{equation*}

We recall that, in the periodic case, $P\Delta u = \Delta u$ if $u \in 
V^2_p$. We shall often use this property in the sequel.

We begin this section by two auxiliary lemmas.
\begin{lemme}\label{estimationv3}
Let $u$ be a weak solution of the Navier-Stokes equations such that
  $w=(I-M)u\in L^\infty(0,T;V^{1/2}_p)\cap L^2(0,T;V^{3/2}_p)$ and
  $v=Mu \in L^\infty(0,T;V_p)\cap L^2(0,T;V^2_p)$. Then we have the
  following estimates,
 for $0 < \gamma \leq {\nu}/(2 \mu^2_0)$ and $0\leq t\leq T$,
 \begin{multline}
  \label{L2v3}
  \nl{v_3(t)}{2}^2\, \leq \, \exp(-\gamma t) \nl{v_3(0)}{2}^2
+\frac{2}{\nu}  \sup_s \nl{A^{-1/2}_\varepsilon (M(Pf(s))_3)}{2}^2\\
+ \frac{2}{\nu}K_7 \varepsilon \exp(-\gamma t)
  \int_0^t  \exp(\gamma s) |w(s)|^2_{1/2} |w(s)|^2_{3/2} ds ~,
\end{multline}
and, for $2 \leq q < +\infty$,
 \begin{multline}
  \label{Lqv3}
   \|v_3(t)\|^2_{L^q} \, \leq \,
 K_8(q) \Big( \|v_3(0)\|^2_{L^q}
 + \varepsilon^{ \frac{2}{q}} \exp(-\gamma t) \int_0^{t} \exp (\gamma s )
 |w(s)|^2_{1/2} |w(s)|^2_{3/2} ds \\
+ \sup_s \|\nabla (-\Delta_2)^{-1}(M(Pf(s))_3)\|^2_{L^q} +
\varepsilon^{-2+  \frac{6}{q}}
\int_{(t-1)^+}^{t} |w(s)|^2_{1/2} |w(s)|^2_{3/2} ds)  \Big) ~,
\end{multline}
where $(t-1)^+ = \sup (0, t-1)$ and $K_8(q)$ is a positive constant
independent of $\varepsilon$, but depending on $q$.
\end{lemme}

\begin{proof}
The function $v_3$ satisfies the following linear equation
\begin{equation}
 \label{v3}
 \partial_t v_3 -\nu\Delta v_3 + \tilde{v} \nabla v_3
 + M((w\nabla w)_3)\,= \,M((Pf)_3)~.
\end{equation}
We first take the scalar product in $L^2(Q_{\varepsilon})$ of the above
equation with $v_3$. Since $\tilde{v}$ and $w$ are divergence-free
vector fields, we obtain, by integrating by parts, that
\begin{equation}
 \label{Deltav3}
 \int_{Q_{\varepsilon}} (\tilde{v}  \nabla v_3)v_3 \, dx\, = \,
 \frac{1}{2} \int_{Q_{\varepsilon}} \tilde{v}  \nabla v_3^2 \, dx\, 
 = \, 0~,
\end{equation}
and that
\begin{equation}
 \label{w3v3}
 \int_{Q_{\varepsilon}} (w  \nabla w_3)v_3 \, dx \, = \,
 - \int_{Q_{\varepsilon}} (w \nabla v_3) w_3 \, dx ~.
\end{equation}
Applying the estimate (\ref{doi}) of the lemma \ref{lema1}  to
the term $ |\int_{Q_{\varepsilon}} (w \nabla v_3) w_3dx |$, we
get, for $0 \leq t \leq T$,
\begin{equation*}
  \frac{1}{2} \partial_t \nl{v_3(t)}{2}^2+ \frac{\nu}{2} |v_3(t)|_1^2\,
  \leq \, \frac{1}{\nu} \nl{A^{-1/2}_\varepsilon(M(Pf(t))_3)}{2}^2 + 
  \frac{1}{\nu}C_1
  \varepsilon^{-1}|w(t)|^4_{1/2}~,
\end{equation*}
or also, by (\ref{Poincare}),
\begin{multline}
\label{2v31}
 \partial_t \nl{v_3(t)}{2}^2+ \frac{\nu}{2} |v_3(t)|_1^2 +
 \frac{\nu}{2 \mu^2_0} \nl{v_3(t)}{2}^2 \, \\
  \leq \, \frac{2}{\nu} \nl{A^{-1/2}_\varepsilon(M(Pf(t))_3)}{2}^2 
+ \frac{2}{\nu}C_1K_0^2 \varepsilon |w(t)|^2_{1/2} |w(t)|^2_{3/2}~.
\end{multline}
Integrating the inequality (\ref{2v31}) and using the Gronwall
lemma, we obtain, for $0 < \gamma \leq {\nu}/(2 \mu^2_0)$ and for
$0 \leq t \leq T$,
\begin{multline}
\label{2v32}
\nl{v_3(t)}{2}^2 + \frac{\nu}{2} \exp(-\gamma t)
\int_0^t  \exp(\gamma s) |v_3(s)|^2_1ds
\, \leq \, \exp(-\gamma t) \nl{v_3(0)}{2}^2 \\
+ \frac{2}{\nu \gamma}  
\sup_s \nl{A^{-1/2}_\varepsilon(M(Pf(s))_3)}{2}^2 \\
+ \frac{2}{\nu}C_1  K_0^2 \varepsilon \exp(-\gamma t)
  \int_0^t  \exp(\gamma s) |w(s)|^2_{1/2} |w(s)|^2_{3/2} ds ~.
\end{multline}
Integrating now the inequality (\ref{2v31}) from $t_0$ to $t_1$, we
deduce from (\ref{2v32}) that, for $0 \leq t_0 \leq t_1$,
\begin{multline}
\label{2v33}
 \int_{t_0}^{t_1} |v_3(s)|_1^2 ds \, \leq \,
 \frac{2}{\nu} \exp(-\gamma t_0) \nl{v_3(0)}{2}^2
+ \frac{4}{\nu^2} ( \frac{1}{\gamma} + t_1-t_0)  \sup_s
\nl{A^{-1/2}_\varepsilon(M(Pf)_3(s))}{2}^2
\\
+  \frac{4}{\nu^2}C_1  K_0^2 \varepsilon \Big( \exp(-\gamma t)
  \int_0^{t_0}  \exp(\gamma s) |w(s)|^2_{1/2} |w(s)|^2_{3/2} ds
 \\
 +
\int_{t_0}^{t_1} |w(s)|^2_{1/2} |w(s)|^2_{3/2} ds \Big)~.
\end{multline}

We now fix a real number $q\geq 2$. Multiplying (\ref{v3}) by
$|v_3|^{q-2}
v_3$, integrating over $Q_{\varepsilon}$ and remarking, as in
(\ref{Deltav3}), that
$\int_{Q_{\varepsilon}} ( \tilde{v}  \nabla v_3 )|v_3|^{q-2}v_3 dx= 0$
we obtain, for $0 \leq t \leq T$
\begin{multline}
\label{egaliteq}
\frac{1}{q} \partial_t \|v_3\|^q_{L^q}+ \nu (q-1)
\int_{Q_{\varepsilon}} |v_3|^{q-2} ( (\partial_{x_1}v_3)^2 +
(\partial_{x_2}v_3)^2 ) dx \\
+ \int_{Q_{\varepsilon}} (w \nabla w_3) |v_3|^{q-2}v_3 dx \, = \,
\int_{Q_{\varepsilon}} (M(Pf)_3) |v_3|^{q-2}v_3 dx~.
\end{multline}
Arguing as in (\ref{w3v3}), we remark that
\begin{equation}
 \label{w3v3q}
 \int_{Q_{\varepsilon}} (w  \nabla w_3) |v_3|^{q-2} v_3 \, dx \, = \,
 - (q-1) \int_{Q_{\varepsilon}} w_3 |v_3|^{q-2} (w_1 \partial_{x_1}v_3
 + w_2 \partial_{x_2}v_3) \, dx ~.
\end{equation}
Furthermore, we have
\begin{equation}
 \label{fv3q}
\int_{Q_{\varepsilon}} (M(Pf)_3) |v_3|^{q-2}v_3 \, dx \, = \,
(q-1)\int_{Q_{\varepsilon}} |v_3|^{q-2} 
\nabla ((-\Delta_2)^{-1} (M(Pf)_3)) \nabla v_3 \, dx~.
\end{equation}
Using H\"{o}lder inequalities, we deduce from the equalities
(\ref{egaliteq}),
(\ref{w3v3q}) and (\ref{fv3q}) that, for $0 \leq t \leq T$,
\begin{multline*}
\frac{1}{q} \partial_t \|v_3(t)\|^q_{L^q}+ \frac{\nu}{2} (q-1)
\int_{Q_{\varepsilon}} |v_3(t)|^{q-2} ( (\partial_{x_1}v_3(t))^2 +
(\partial_{x_2}v_3(t))^2 ) dx \, \leq\, \\
(q-1)\nu^{-1} \Big(
\int_{Q_{\varepsilon}} |v_3(t)|^{q-2}
|\nabla ((-\Delta_2)^{-1}(M(Pf(t))_3))|^2 dx \\
+ \varepsilon^{-2 + 2/q} \sum_{i=1}^{2} \|v_3(t)\|^{q-2}_{L^q}
\|w_3(t)\|^2_{L^{2q,2}} \|w_i(t)\|^2_{L^{2q,2}} \Big)~,
\end{multline*}
or also, due to Lemma \ref{lema2}
\begin{equation}
\label{estv3q1}
\frac{1}{q} \partial_t \|v_3(t)\|^q_{L^q}\, \leq\,
(q-1)\nu^{-1} C_2 \|v_3(t)\|^{q-2}_{L^q} \big(
\|\nabla (-\Delta_2)^{-1}(M(Pf(t))_3) \|^2_{L^q} + 
\varepsilon^{-2 + \frac{2}{q}} |w(t)|^4_{1-\frac{1}{q}} \big)~.
\end{equation}
Since $|w|_{1-1/q} \leq C_3 |w|^{1/2 +1/q}_{1/2} |w|^{1/2 -1/q}_{3/2}$,
we derive from (\ref{estv3q1}) and the Poincar\'{e} inequality for
$w$  that
\begin{equation}
\label{estv3q2}
\frac{1}{2} \partial_t \|v_3(t)\|^2_{L^q}\, \leq\,
(q-1)\nu^{-1} C_4 \big(
\|\nabla (-\Delta_2)^{-1}(M(Pf(t))_3) \|^2_{L^q} + \varepsilon^{-2 + 6/q}
|w(t)|^2_{1/2}|w(t)|^2_{3/2}\big)~.
\end{equation}
Integrating the estimate (\ref{estv3q2}), we obtain, for $0\leq t\leq
\inf(1,T)$,
\begin{multline}
\label{estv3q3}
 \|v_3(t)\|^2_{L^q}\, \leq\,  \|v_3(0)\|^2_{L^q} \\
+ (q-1)\nu^{-1} C_4 \big( \sup_s \|\nabla (-\Delta_2)^{-1}
(M(Pf(s))_3) \|^2_{L^q} + \varepsilon^{-2 + 6/q}
\int_{0}^{t}|w(s)|^2_{1/2}|w(s)|^2_{3/2} ds \big)~.
\end{multline}
Using the uniform Gronwall lemma, we also deduce from (\ref{estv3q2})
that, for $1 \leq t \leq T$,
\begin{multline}
\label{Gronv3q}
 \|v_3(t)\|^2_{L^q}\, \leq\, \int_{t-1}^{t} \|v_3(s)\|^2_{L^q} ds  
+(q-1)\nu^{-1} C_4 \big(
\sup_s \|\nabla (-\Delta_2)^{-1} (M(Pf(s))_3) \|^2_{L^q} \\
+ \varepsilon^{-2 + 6/q}
\int_{t-1}^{t}|w(s)|^2_{1/2}|w(s)|^2_{3/2} ds \big)~.
\end{multline}
But, from the Sobolev embedding $H^1(\Omega) \subset  L^q(\Omega)$, for
$1\leq q< +\infty$ and the inequality (\ref{2v33}), we infer that
\begin{multline}
\label{Gronv3q2}
\int_{t-1}^{t} \|v_3(s)\|^2_{L^q} ds \, \leq \,
\varepsilon^{- 1 + 2/q} \int_{t-1}^{t} |v_3(s)|^2_1 ds \, \\
\leq \,  C_q \Big( \|v_3(0)\|^2_{L^q}
+  \sup_s \|\nabla (-\Delta_2)^{-1} (M(Pf(s))_3)\|^2_{L^q} + \\
+ \varepsilon^{2/q} \big[ \exp(-\gamma t) \int_0^{t} \exp(\gamma s)
|w(s)|^2_{1/2} |w(s)|^2_{3/2} ds + \int_{t-1}^{t} |w(s)|^2_{1/2}
|w(s)|^2_{3/2} ds \big] \Big) ~.
\end{multline}
Finally the estimates (\ref{estv3q3}), (\ref{Gronv3q}) and
(\ref{Gronv3q2}) imply that, for $0 \leq t \leq T$,
\begin{multline}
\label{v3qfinal}
 \|v_3(t)\|^2_{L^q} \, \leq \,
 \tilde{C}_q \Big( \|v_3(0)\|^2_{L^q}
 + \varepsilon^{\frac{2}{q}} \exp(-\gamma t) \int_0^{t} \exp (\gamma s )
 |w(s)|^2_{1/2} |w(s)|^2_{3/2} ds \\
+ (q-1) \big(\sup_s \|\nabla (-\Delta_2)^{-1} (M(Pf(s))_3)\|^2_{L^q} +
\varepsilon^{-2+ \frac{6}{q}}
\int_{(t-1)^+}^{t} |w(s)|^2_{1/2} |w(s)|^2_{3/2} ds \big) \Big) ~,
\end{multline}
where $(t-1)^+ = \sup (0, t-1)$.
\end{proof}

\begin{lemme}\label{estimationH1v}
Let $u$ be a weak solution of the Navier-Stokes equations such that
  $w=(I-M)u\in L^\infty(0,T;V^{1/2}_p)\cap L^2(0,T;V^{3/2}_p)$ and
  $v=Mu \in L^\infty(0,T;V_p)\cap L^2(0,T;V^2_p)$. Then we have the
  following estimate, for $0< \gamma \leq \nu \mu_0^{-2}$ and
 $0\leq t\leq  T$,
 \begin{multline}
  \label{H1v}
|\tilde{v}(t)|_1^2 \, \leq \, |\tilde{v}(0)|_1^2
+  K_9 \Big( \sup_s \nl{M \widetilde{Pf}(s)}{2}^2  \\
+\varepsilon^{-1} \exp(-\gamma t) \int_0^{t} \exp(\gamma s)
|w(s)|_{1/2}^2 |w(s)|_{3/2}^2 ds
\Big)~.
 \end{multline}
\end{lemme}
\begin{proof}
We first recall the equation satisfied by $\tilde{v}$, that is,
\begin{equation*}
 \partial_t \tilde{v} -\nu\Delta \tilde{v} + 
M (\widetilde{B_{\varepsilon}(v,v)})
 + M (\widetilde{B_{\varepsilon}(w,w)})\,= \,M(\widetilde{Pf})~.
\end{equation*}
Taking the scalar product in $L^2(Q_{\varepsilon})$ of the above
equation with $- \Delta \tilde{v}$ and remarking, as in
\cite{RaugelSell94}  that
\begin{equation*}
 \int_{Q_{\varepsilon}} (v  \nabla v)\Delta \tilde{v} \, dx\,=
\int_{Q_{\varepsilon}} (\tilde{v}  \nabla \tilde{v})\Delta 
\tilde{v} \, dx
\,= \, \varepsilon \int_{\Omega} (\tilde{v}   \nabla \tilde{v})
\Delta \tilde{v} \, dx_1dx_2 \, =\, 0~,
\end{equation*}
we obtain the equality
\begin{equation}
\label{f0v}
  \frac{1}{2} \partial_t |\tilde{v}|_1^2 +\nu|\tilde{v}|_2^2 +
  \int_{Q_{\varepsilon}}  ( \widetilde{w \nabla w}) (- \Delta \tilde{v}) 
  \, dx
 \, =\, \int_{Q_{\varepsilon}} \widetilde{Pf} (- \Delta \tilde{v}) \, dx~.
 \end{equation}
 Since, by the estimate (\ref{doi}) of Lemma \ref{lema1},
\begin{equation*}
 \int_{Q_{\varepsilon}} ( \widetilde{w \nabla w}) (- \Delta
\tilde{v}) dx \, \leq \, C \varepsilon^{-1/2} |w|_{1/2} |w|_{3/2}
\nl{\Delta \tilde{v}}{2}~,
\end{equation*}
 we infer from (\ref{f0v}), by using also a Young inequality, that,
 for $0 \leq t \leq T$,
\begin{equation}
\label{f1v}
 \partial_t |\tilde{v}(t)|_1^2 + \nu|\tilde{v}(t)|_2^2 \, \leq \,
2 {\nu}^{-1} \nl{M \widetilde{Pf}(t)}{2}^2 + 2 {\nu}^{-1} C^2
\varepsilon^{-1}
|w(t)|_{1/2}^2 |w(t)|_{3/2}^2~,
\end{equation}
or also
\begin{equation}
\label{f1vbis}
 \partial_t |\tilde{v}(t)|_1^2 +
\frac{\nu}{\mu^2_0}|\tilde{v}(t)|_1^2 \, \leq \,
2 {\nu}^{-1} \nl{M \widetilde{Pf}(t)}{2}^2 + 2 {\nu}^{-1} C^2
\varepsilon^{-1}
|w(t)|_{1/2}^2 |w(t)|_{3/2}^2~,
\end{equation}
Integrating the inequality (\ref{f1vbis}) and using the Gronwall
lemma, we obtain, for $0 < \gamma \leq \nu \mu^{-2}_0$ and
for $0 \leq t \leq T$,
\begin{multline}
\label{f2v}
|\tilde{v}(t)|_1^2 \, \leq \,  \exp(-\gamma t)|\tilde{v}(0)|_1^2 +
\frac{2}{\nu \gamma} \sup_s \nl{M \widetilde{Pf}(s)}{2}^2\\
 + 2 {\nu}^{-1} C^2 \varepsilon^{-1}  \exp(-\gamma t)\int_{0}^{t}
 \exp(\gamma s)|w(s)|_{1/2}^2 |w(s)|_{3/2}^2 ds~,
\end{multline}
which at once implies the estimate (\ref{H1v}) of Lemma
\ref{estimationH1v}.
\end{proof}

Now we can prove Theorem \ref{theoreme2}.

\begin{proof}
Like in the proof of Theorem \ref{theoreme1}, we consider a Galerkin
approximation, using the first $m$ eigenfunctions $\psi_1$,
$\psi_2$, ....,$\psi_m$ of the Stokes
operator $A_{\varepsilon}$. As in Theorem \ref{theoreme1}, these
eigenfunctions $\psi_j$ are chosen so that, either $\psi_j \in
MH_p$ or $\psi_j \in (I-M)H_p$. Moreover, if
the eigenvector $\psi_j$ is independent of the third variable $x_3$,
it can be chosen so that, either $\psi_j \equiv  M\psi_j =
(M\psi_{j1}, M\psi_{j2}, 0)$ or $\psi_j = (0,0, M\psi_{j3})$. These
properties imply that,
if $\mathcal{P}_m : H_p \rightarrow H_p$ denotes the projector
onto the space $\mathcal{V}_m$ generated
by the first $m$ eigenfunctions, then, for every $s \in [0,2]$ and
for every $u \in V^s_p$, the inequalities
\begin{equation}
\label{5Pm}
|\mathcal{P}_m Mu_3|_s \, \leq \, |Mu_3|_s~, \quad
|\mathcal{P}_m \widetilde{Mu}|_s\, \leq \, |\widetilde{Mu}|_s~,
\end{equation}
as well as the inequalities (\ref{4Pm}) hold.
We recall that $\mathcal{P}_m(I-M)u_0$ (resp.
$\mathcal{P}_mMu_0$) converges to $(I-M)u_0$ (resp. $Mu_0$) in
$V^{1/2}_p$ (resp. $V_p$), as $m$ goes to $+ \infty$.

 Like in the proof of Theorem \ref{theoreme1}, we know 
(see \cite{ConstantinFoias}, Chapter 8, for example, or \cite{Temam84}) that,
for every $m \in \mathbb{N}$, there exists a global solution 
$u_m \in C^1([0,+\infty); V^2_p \cap \mathcal{V}_m)$ of the equations
(\ref{NSabs}) or also of (\ref{NSabsv}) and
(\ref{NSabsw}), where $B_\varepsilon$ is replaced by $\mathcal{P}_m
B_\varepsilon$ and $Pf$ by $\mathcal{P}_m Pf$ and where the initial
condition is $u_m(0)=\mathcal{P}_m (I-M)u_0 + 
\mathcal{P}_m Mu_0 \equiv w_{0m} + v_{0m}$. 
Moreover, for every $\tau >0$, $u_m$ and $\partial_t u_m$ are uniformly bounded 
with respect to $m$ in the spaces $L^{\infty}(0, +\infty ;H_p) \cap L^2(0, \tau;V_p)$
and $L^{4/3}(0, \tau;V'_p)$ respectively. We want to show that
this solution $u_m$ satisfies the additional estimates and properties
given in Theorem \ref{theoreme2}. In order to simplify
the notation, we drop the subscript $m$, when there is no confusion.
Like in the proof of Theorem \ref{theoreme1}, we take
the scalar product in $L^2(Q_{\varepsilon})$
 of the modified equation (\ref{NSabsw}) with
$A^{1/2}w=(-\Delta)^{1/2}w$ and obtain the equality
(\ref{ff0}).
Applying the inequality (\ref{unu}) of Lemma \ref{lema1}, we have,
for $t \geq 0$,
\begin{equation}
  \label{f1w}
\int_{Q_\varepsilon} (w\nabla w (-\Delta)^{\frac{1}{2}}w)(t,x) )dx
  \leq \, K_1|w(t)|_{\frac{1}{2}}
 |\nabla w(t)|_{\frac{1}{2}} |(-\Delta)^{\frac{1}{2}}w(t)|_{\frac{1}{2}}
 \leq  \,  C|w(t)|_{1/2} |w(t)|_{3/2}^2~.
\end{equation}
In order to estimate the term $\int_{Q_\varepsilon} v\nabla w
((-\Delta)^{1/2}w) dx$, we apply Lemma \ref{estimation2d} and
obtain, for $t \geq 0$,
\begin{equation}
  \label{f2w}
  \int_{Q_\varepsilon} (v\nabla w (-\Delta)^{1/2}w)(t,x) dx
\,  \leq  \, K_5 \varepsilon^{-1/2} |v(t)|_1 |w(t)|_{1/2} |w(t)|_{3/2}
\,  \leq  \, C\varepsilon^{1/2} |v(t)|_1 |w(t)|_{3/2}^2~.
\end{equation}
To estimate the third nonlinear term
$\int_{Q_\varepsilon} w\nabla v ((-\Delta)^{1/2}w) dx$,
 we can use the estimate (\ref{doi}) of Lemma \ref{lema1} as follows,
\begin{equation}
  \label{f3w}
  \begin{split}
\int_{Q_\varepsilon} (w\nabla v (-\Delta)^{1/2}w)(t,x) dx
  &\leq K_2\varepsilon^{-1/2}|\nabla v(t)|_0 |w(t)|_{1/2}
  |(-\Delta)^{1/2}w(t)|_{1/2}\\
  &\leq C\varepsilon^{1/2} |v(t)|_1 |w(t)|_{3/2}^2~.
   \end{split}
\end{equation}

Finally, like in the proof of Theorem \ref{theoreme1}, we
write, for $t \geq 0$,
\begin{equation}
  \label{f4w}
\int_{Q_\varepsilon}( (I-M)f(-\Delta)^{1/2}w)(t,x) dx 
\, \leq \, C \nu^{-1}\varepsilon \nl{(I-M)Pf(t)}{2}^2 +
  \frac{\nu}{4}|w(t)|_{3/2}^2~.
\end{equation}

Due to the estimates (\ref{ff0}), (\ref{f1w}), (\ref{f2w}), (\ref{f3w})
and (\ref{f4w}), we have, for $t \geq 0$,
\begin{equation}
\label{inegresum}
  \partial_t |w(t)|_{1/2}^2 + 2 (\frac{3\nu}{4} - C_1|w(t)|_{1/2} -
 C_2 \varepsilon^{1/2} |v(t)|_1 )
  |w(t)|_{3/2}^2 \leq
C_3\nu^{-1} \varepsilon \nl{(I-M)Pf(t)}{2}^2.
\end{equation}
Due to the property (\ref{4Pm}) and
to the hypothesis (\ref{H1th}) on the initial
conditions, where $k_1$, $k_2$ are small enough, there exists a
positive time $T$ such that, for $t \in [0,T)$,
\begin{equation}
  \label{Hwv}
 C_1|w(t)|_{1/2} + C_2 \varepsilon^{1/2} |v(t)|_1 \,< \, \frac{\nu}{2}~.
\end{equation}
and, that, if $T < \infty$,
\begin{equation}
  \label{Hwvegal}
 C_1|w(T)|_{1/2} + C_2 \varepsilon^{1/2} |v(T)|_1 \, =\, \frac{\nu}{2}~.
\end{equation}
We shall show by contradiction that $T=+\infty$. To this end, we
shall estimate separately the terms $|w(t)|_{1/2}$, $|\tilde{v}(t)|_1$
and $|v_3(t)|_1$. The estimate of the term $|\tilde{v}(t)|_1$ will be
a consequence of Lemma \ref{estimationH1v}.

We derive from the estimates (\ref{inegresum}), (\ref{Hwv}) and
(\ref{Hwvegal}) that, for
$t \in [0,T]$,
\begin{equation}
  \label{f5w}
  \partial_t |w(t)|_{1/2}^2 +\frac{\nu}{2}|w(t)|_{3/2}^2 \, \leq\,
C_3\nu^{-1}\varepsilon\nl{(I-M)Pf(t)}{2}^2,
\end{equation}
 which in turn implies that
 \begin{equation}
 \label{f6w}
 \begin{split}
   \partial_t |w(t)|_{1/2}^2 +\frac{\nu}{2} \varepsilon^{-2}
  K_0^{-2}|w(t)|_{1/2}^2 \, \leq \, 
&  \partial_t |w(t)|_{1/2}^2 +\frac{\nu}{4} \varepsilon^{-2}
  K_0^{-2}|w(t)|_{1/2}^2
 + \frac{\nu}{4} |w(t)|_{3/2}^2  \\
\leq\,&
C_3\nu^{-1}\varepsilon\nl{(I-M)Pf(t)}{2}^2~.
\end{split}
\end{equation}
The Gronwall lemma then gives, for $t \in [0,T]$,
\begin{equation}
  \label{est1w}
  |w(t)|_{1/2}^2 \, \leq \, \exp(- \frac{\nu}{2} \varepsilon^{-2}
  K_0^{-2} t )|w_0|_{1/2}^2
  + C_4\nu^{-2} \varepsilon^3 \sup_s \nl{(I-M)Pf(s)}{2}^2~.
\end{equation}
On the other hand, integrating the inequalities (\ref{f6w}),
we get, for $0< \gamma \leq \frac{\nu}{4}
\varepsilon^{-2} K_0^{-2}$ and for $0 \leq t_1< t_2 \leq  T$,
\begin{multline}
\label{est2w}
|w(t_2)|_{1/2}^2 +\exp(-\gamma t_2) \frac{\nu}{4} \int_{t_1}^{t_2}
\exp(\gamma s) |w(s)|_{3/2}^2  d{s} \,\leq \,
\exp(-\gamma (t_2 - t_1) ) |w(t_1)|_{1/2}^2 \\
+C_3 (\gamma \nu)^{-1}(1- \exp(-\gamma (t_2-t_1))) \varepsilon \sup_s
\nl{(I-M)Pf(s)}{2}^2~ .
\end{multline}
We now fix a positive
number $\gamma$, satisfying $0 < \gamma \leq \inf(
\frac{\nu}{2\mu^2_0}, \frac{\nu}{4} \varepsilon^{-2} K_0^{-2}) $.
We deduce from the estimates (\ref{est1w}) and (\ref{est2w}) that,
 for $t \in [0,T]$,
 \begin{equation}
 \label{winte0t}
 \begin{split}
\exp(-\gamma t) \int_{0}^{t}
\exp(\gamma s) |w(s)|_{1/2}^2 &|w(s)|_{3/2}^2 d{s}  \,\leq \,
|w_0|^4_{\frac{1}{2}} + C_3 C_4 \gamma^{-1}\nu^{-3} \varepsilon^4
\sup_s \nl{(I-M)Pf(s)}{2}^4 \\
& \phantom{|w(s)|_{3/2}^2 d{s}  \leq }
+ |w_0|^2_{\frac{1}{2}} \sup_s \nl{(I-M)Pf(s)}{2}^2 
[C_4 \nu^{-2}\varepsilon^3 \\
&\phantom{|w(s)|_{3/2}^2 d{s}  \leq }
+ C_3(\gamma \nu)^{-1} \varepsilon (1- \exp(-\gamma t))
\exp(- \frac{\nu}{2} \varepsilon^{-2} K_0^{-2} t )  ] \\
 & \quad\leq \, C_5|w_0|^4_{\frac{1}{2}}
 + C_6 \sup_s \nl{(I-M)Pf(s)}{2}^4 ( \varepsilon^4 +
 \varepsilon^2 E(t))~,
\end{split}
\end{equation}
where $E(t) = (1- \exp(-\gamma t))^2
 \exp(- \nu \varepsilon^{-2} K_0^{-2} t )$.
On the one hand, we remark that, for $t \geq t_\varepsilon$,
where
$t_\varepsilon = - 2 \varepsilon^2 K^2_0~\nu^{-1} \ln \varepsilon$,
$E(t) \leq \varepsilon^2$. On the other hand, for  $t \leq
t_{\varepsilon}$, we notice that $E(t) \leq (1- \exp(-\gamma t))^2
\leq \gamma^2 t^2_{\varepsilon} \leq C \varepsilon^2$. From these
remarks and from the estimate (\ref{winte0t}), we
finally infer that, for $t \in [0,T]$,
\begin{equation}
\label{intwfinal}
\exp(-\gamma t) \int_{0}^{t}
\exp(\gamma s) |w(s)|_{1/2}^2 |w(s)|_{3/2}^2 d{s} \,\leq \,
\varepsilon D_1~,
\end{equation}
where
$D_1=C_7 (\varepsilon^{-1} |w_0|_{1/2}^4
 + \varepsilon^3 \sup_s \nl{(I -M) Pf(s)}{2}^4)$.

Lemma \ref{estimationH1v}, the inequality (\ref{intwfinal}) and
the property (\ref{4Pm}) imply that,
for $t \in [0,T]$,
\begin{equation}
\label{f4v}
|\tilde{v}(t)|_1^2 \, \leq \, D_0 + D_1~,
\end{equation}
where $D_0 = |\tilde{v}_0|_1^2 +
C_8 \sup_s \nl{M(\widetilde{Pf})(s)}{2}^2$.

It remains to estimate the term $|v_3(t)|_1$.
Taking the scalar product in $L^2(Q_{\varepsilon})$ of the modified
equation (\ref{v3}) with $A_\varepsilon v_3$, applying the estimate
(\ref{doi}) of Lemma  \ref{lema1} as well as the estimate
(\ref{com5}) of Lemma  \ref{lemacinq}, we obtain, for $t \in
[0,T]$,
\begin{multline*}
   \partial_t |v_3(t)|_1^2 + 2\nu |v_3(t)|_2^2\, 
  \leq \, 2 |v_3(t)|_2  \bigl( \nl{\mathcal{P}_mM(Pf)_3(t)}{2}\\
+  K_2 \varepsilon^{-1/2} |w(t)|_{1/2} |w(t)|_{3/2}
+ K_6 \varepsilon^{-1/2} |\tilde{v}(t)|_1 |v_3(t)|_1  \bigr)~,
\end{multline*}
or also
\begin{multline}
\label{v31}
   \partial_t |v_3(t)|_1^2 + \frac{2\nu}{\mu^2_0}|v_3(t)|_1^2 \, \leq
   \, \partial_t |v_3(t)|_1^2 + 2\nu |v_3(t)|_2^2\, \leq \, \\
   \leq \,  C_9 \nu^{-1} \big( \nl{\mathcal{P}_mM(Pf)_3(t)}{2}^2
+ \varepsilon^{-1}  |w(t)|_{1/2}^2 |w(t)|_{3/2}^2
+ \varepsilon^{-1} |\tilde{v}(t)|_1^2 |v_3(t)|_1^2 \big)~.
\end{multline}
By integration, it follows from (\ref{v31}) and (\ref{5Pm}) that,
for $t \in [0,T]$,
\begin{multline}
\label{v32}
|v_3(t)|_1^2   \, \leq \,  \exp(-\gamma t) |v_3(0)|_1^2
+ C_9\nu^{-1} \Big( \gamma^{-1} \sup_s \nl{M(Pf)_3(s)}{2}^2 \\
+ \varepsilon^{-1} \exp(-\gamma t)\int_{0}^{t}
\exp(\gamma s) |w(s)|_{1/2}^2 |w(s)|_{3/2}^2 ds \\
+ \varepsilon^{-1} \exp(-\gamma t)\sup_s |\tilde{v}(s)|_1^2
\int_{0}^{t} \exp(\gamma s) |v_3(s)|_1^2 ds \Big)~.
\end{multline}
We infer from (\ref{2v32}), (\ref{est2w}), (\ref{intwfinal}) and
(\ref{v32}) that, for $t \in [0,T]$,
\begin{equation}
\label{v33}
|v_3(t)|_1^2 \, \leq \,  |v_3(0)|_1^2 + C_{10} \Big( \sup_s \nl{M(Pf)_3(s)}{2}^2
+ D_1 +  (D_0 + D_1)D_2 \Big)~,
\end{equation}
where
\begin{equation}
\label{D2}
D_2 \, = \, \varepsilon^{-1} \big( \nl{v_3(0)}{2}^2
+ \sup_s \nl{A^{-1/2}_\varepsilon (M(Pf)_3(s))}{2}^2 \big) + \varepsilon D_1~.
\end{equation}
Finally, the inequalities (\ref{f4v}) and (\ref{v33}) give, for $t
\in [0,T]$,
\begin{equation}
\label{vfinal}
|v(t)|_1^2 \, \leq \,  |v_3(0)|_1^2 + D_0 +D_1
+ C_{10} \Big( \sup_s \nl{M(Pf)_3(s)}{2}^2
+ D_1 +  (D_0 + D_1)D_2 \Big) ~.
\end {equation}
If $k_1$, $k_2$, $k_3$, $k_4$, $k_5$ and
$k_6$ are small enough, the properties (\ref{4Pm}), (\ref{5Pm}),
the hypotheses (\ref{H1th}) and (\ref{H2th})
together with the estimates (\ref{est1w}) and  (\ref{vfinal}) imply
that, for $t \in [0,T]$,
\begin{equation}
\label{contra}
 C_1|w(T)|_{1/2} + C_2 \varepsilon^{1/2} |v(T)|_1 \, <\,
 \frac{\nu}{4}~,
\end{equation}
which contradicts the equality (\ref{Hwvegal}). It follows that
$T=+\infty$.

We have just proved that, under the hypotheses
(\ref{H1th}) and (\ref{H2th}), for any $m \in \mathbb{N}$, the
solution $u_m \in C^1([0, +\infty); \mathcal{V}_m)$ of the modified
Navier-Stokes equations (\ref{NSabs}) with initial data
$u_m(0)= \mathcal{P}_m u_{0m}$ satisfies
\begin{equation}
\label{5inegam1}
\sup_{t \geq 0} (|w_m(t)|_{1/2} + \varepsilon^{1/2} |v_m(t)|_1) \, <\,
C_{11}~,
\end{equation}
where $C_{11}$ is a positive constant independent of $\varepsilon$
and $m$.
Integrating the inequalities (\ref{f5w}), (\ref{f1v}) and
(\ref{v31}) and using the estimates (\ref{5inegam1}), (\ref{f4v}),
(\ref{2v31}) as well as the hypotheses (\ref{H1th}) and (\ref{H2th}),
one also shows that, for any $t \in [0,+ \infty)$,
\begin{equation}
\label{5inegam2}
\int_{0}^{t} (|w_m(s)|^2_{3/2} + \varepsilon |v_m(s)|^2_2) d{s} \,
\leq \, \varepsilon^{-1} C_{12} t~,
\end{equation}
where $C_{12}$ is a positive constant independent of $\varepsilon$
and $m$.

Like in the proof of Theorem
\ref{theoreme1}, a classical argument (see \cite{ConstantinFoias},
Chapter 8 or \cite{Temam84}) together with the estimates 
(\ref{5inegam1}) and (\ref{5inegam2}),
shows that $u =\lim_{m \rightarrow
+\infty} u_m$ belongs to the space $L^{\infty}(0,\infty; V^{1/2}_p)
\cap L^2_{loc}([0,\infty); V^{3/2}_p)$, is a weak Leray solution of
the equations (\ref{NSabs}) with initial data $u(0)=u_0$ and that 
$Mu \in L^\infty(0,\infty;V_p)\cap L^2_{loc}([0,\infty);V^2_p)$. The
uniqueness of the solution $u$ follows from Theorem \ref{unic}.
Arguing as in Remark \ref{rem}, we actually show that
$\partial_t u$ belongs to
$L^2_{loc}([0,\infty); V^{-1/2}_p)$. Indeed, we deduce from the 
equality (\ref{3uphi}), Lemma
\ref{lema1} and Remark \ref{remarque} that, for any $t \geq 0$,
for any $\varphi \in L^2(0,t; V^{1/2}_p)$, 
\begin{equation*}
\int_{0}^{t}\int_{Q_{\varepsilon}} (u\nabla u)(s,x) \varphi(s,x) \, 
dx ds \, \leq \, C_\varepsilon 
\|u\|_{L^{\infty}(0,t; V^{1/2}_p)} \|u\|_{L^2(0,t; V^{3/2}_p)}
\|\varphi \|_{L^2(0,t; V^{1/2}_p)}~,
\end{equation*}
which implies that $u\nabla u $ belongs to
$L^2_{loc}([0,\infty); V^{-1/2}_p)$. It follows, since $\Delta u$
and $Pf$ also
belong to this space, that  $\partial_t u$ belongs to
$L^2_{loc}([0,\infty); V^{-1/2}_p)$. As $u \in
L^2_{loc}([0,\infty); V^{3/2}_p) \cap H^1_{loc}([0,\infty);
V^{-1/2}_p)$, $u$ is also in the space
$C^0([0,+\infty); V^{1/2}_p)$. The vector $v=Mu$ actually lies in the 
space $C^0([0,+ \infty);V_p)$. Indeed, applying the estimate 
(\ref{doi}) of Lemma \ref{lema1} and Remark \ref{remarque}, 
we obtain, for $t\geq 0$ and $\varphi \in L^2(0,t; H_p)$,
\begin{multline*}
\int_{0}^{t}\int_{Q_{\varepsilon}} (v\nabla v +w\nabla w)(s,x)
M\varphi(x)\, dx  ds \, \\
\leq  C_\varepsilon (
\|v\|_{L^{\infty}(0,t; V^{1/2}_p)}\|v\|_{L^2(0,t; V^{3/2}_p)}
+ \|w(s)\|_{L^{\infty}(0,t; V^{1/2}_p)} \|w(s)\|_{L^2(0,t; 
V^{3/2}_p)}) \| \varphi \|_{L^2(0,t; H_p)}~,
\end{multline*}
 which implies that $MB_\varepsilon(v,v) + MB_\varepsilon(w,w)$
belongs to the space $L^2_{loc}([0,\infty); H_p)$. As
$v \in L^2_{loc}([0,\infty); V^2_p)$ and
$Pf \in L^2_{loc}([0,\infty); H_p)$, we deduce
from the equation (\ref{NSabsv}) that $\partial_t v
\in L^2_{loc}([0,\infty); H_p)$ and thus that $v \in C^0([0,+ \infty);
V_p)$.
\end{proof}

In some sense, we can improve the global
existence results given in Theorem \ref{theoreme2},
if, in the various estimates, we also take into account the $L^q$-norm
of $v_3$, where, for instance, $q \geq 3$.  The hypotheses
in the following theorem are rather involved, but, in the
applications, it allows to take larger initial data and forcing terms.

\begin{theoreme} \label{theoreme3}
Let $\varepsilon_0>0$ be fixed. For any real number $q > 2$,
there exist positive constants $k_1(q)$, $k_2(q)$,
$k_3(q)$, $k_4(q)$, $k_5(q)$ and $k_6(q)$ such that,
for $0< \varepsilon \leq \varepsilon_{0}$, if the initial data
 $(Mu_0, (I-M)u_0) \in V_p \times V^{1/2}_p$ and the force $f\in
L^\infty(0,\infty;(L^2(Q_\varepsilon))^3)$ satisfy
\begin{equation}
  \label{H1thq}
 \begin{split}
  &|Mu_0|_1 \, \leq \, k_1(q)\varepsilon^{-1/2},
  \quad |(I-M)u_0|_{1/2}  \,\leq \, k_2(q) \\
  &\sup_t \nl{M\widetilde{Pf}(t)}{2}\leq k_3(q) \varepsilon^{-1/2}, \quad
  \sup_t \nl{(I-M)Pf(t)}{2}\, \leq \, k_4(q) \varepsilon^{-1}\\
  &\|Mu_{03}\|_{L^q} +  \sup_t \|\nabla (-\Delta_2)^{-1}(MPf_3)(t)
  \|_{L^q} \, \leq \,  k_5(q) \varepsilon^{-1 + 3/q}~,
\end{split}
\end{equation}
and the additional condition
\begin{multline}
\label{H2thq}
\Big( \varepsilon^{1 - 3/q} (\|Mu_{03}\|_{L^q} +
\sup_t \|\nabla (-\Delta_2)^{-1}(MPf_3)(t)\|_{L^q})
+  |(I-M)u_0|^2_{1/2} \\
+ \varepsilon^2 \sup_t\nl{(I-M)Pf(t)}{2}^2 \Big) \times
\Big( \varepsilon^{1/2} (|Mu_{30}|_1 + \sup_t\nl{M(Pf)_3(t)}{2}) +
\mathcal{A}_0
\Big) \, \leq \, k_6(q)~,
\end{multline}
where $\mathcal{A}_0 $ has been defined in Theorem
\ref{theoreme2}, then there exists a global solution
$u(t) \in C^0([0,\infty); V^{1/2}_p) \cap
L^{\infty}(0,\infty; V^{1/2}_p) \cap L^2_{loc}([0,\infty); V^{3/2}_p)$
of (\ref{NSabs}) which is unique in the class of weak Leray solutions.
Moreover, $Mu \in C^0([0,\infty); V_p) \cap
L^\infty(0,\infty;V_p)\cap L^2_{loc}([0,\infty);V^2_p)$
and $u(t)$ satisfies the estimates (\ref{est1w}), (\ref{f4v}),
(\ref{v3qest}) and (\ref{5inegamq}), for every $t\geq 0$.
\end{theoreme}
\begin{proof}
We use the same Galerkin basis as in the proof of Theorem
\ref{theoreme2}, so that the properties (\ref{4Pm}) and (\ref{5Pm})
hold. Since $\mathcal{P}_m Mu_{0}$ converges to $Mu_{0}$ in
$V_p$, $\mathcal{P}_m Mu_{03}$ also converges to $Mu_{03}$ in
$L^q(\Omega)$. Hence, there exists $m_1 =m_1(\varepsilon,q)$ such
that, for $m \geq m_1$,
$\| \mathcal{P}_m Mu_{03} \|_{L^q(\Omega)} \leq 2
\|Mu_{03} \|_{L^q(\Omega)}$ and thus that
\begin{equation}
\label{5Pm3q1}
\| \mathcal{P}_m Mu_{03} \|_{L^q(Q_\varepsilon)} \, \leq \, 2
\|Mu_{03} \|_{L^q(Q_\varepsilon)}~.
\end{equation}
Likewise, for any $t\in [0,+\infty)$,
$\mathcal{P}_m \nabla (-\Delta_2)^{-1} (MPf_3)(t)$ converges to
$\nabla (-\Delta_2)^{-1} (MPf_3)(t)$ in $L^q(\Omega)$, when $m$ goes
to $+\infty$. But, since $\{\nabla (-\Delta_2)^{-1}(MPf)(t)\, | \, t\in
[0,+\infty)\}$ is a bounded set in $V_p$,
$\{ \nabla (-\Delta_2)^{-1}(MPf_3)(t)\, | \, t\in [0,+\infty)\}$ is a
compact set in $L^q(\Omega)$ and thus, there exists
$m_2 =m_2(\varepsilon,q)$ such
that, for $m \geq m_2$, for $t \in [0,+\infty)$,
\begin{equation*}
\| \mathcal{P}_m \nabla (-\Delta_2)^{-1} (MPf_3)(t)
\|_{L^q(\Omega)} \, \leq \, 2
\sup_t \|\nabla (-\Delta_2)^{-1}(MPf_3)(t)\|_{L^q(\Omega)}~,
\end{equation*}
and 
\begin{equation}
\label{5Pm3q2}
\| \mathcal{P}_m \nabla (-\Delta_2)^{-1} (MPf_3)(t)
\|_{L^q(Q_\varepsilon)} \, \leq \, 2
\sup_t \|\nabla (-\Delta_2)^{-1}(MPf_3)(t)\|_{L^q(Q_\varepsilon)}~.
\end{equation}
We set $m_0= \sup(m_1,m_2)$. Like in the proof of Theorem
\ref{theoreme2}, for every $m\geq m_0$, we know that there exists
a global solution $u_m \equiv v_m+w_m=Mu_m + (I-M)u_m$ of
 the equations (\ref{NSabs}),
where $B_\varepsilon$ is replaced by $\mathcal{P}_m
B_\varepsilon$ and $Pf$ by $\mathcal{P}_m Pf$ and where the initial
condition is $u_m(0)=\mathcal{P}_m u_0 \equiv
w_{0m} + v_{0m}$. We shall prove a priori estimates on the
solution $u_m$ . We
again drop the subscript $m$, when there is no confusion. Like in the
proof of Theorem \ref{theoreme2}, taking the inner product in
$L^2(Q_\varepsilon)$ of the modified equation (\ref{NSabsw})
with $A^{1/2}w$, we are led to estimate
$\int_{Q_\varepsilon} w\nabla w ((-\Delta)^{1/2}w )dx$,
$\int_{Q_\varepsilon} v\nabla w ((-\Delta)^{1/2}w) dx $ and
$\int_{Q_\varepsilon} w\nabla v ((-\Delta)^{1/2}w) dx$. The estimate of
the first term does not change and is given in (\ref{f1w}).
Decomposing $v$ into $\tilde{v} + v_3$ and applying the inequality
(\ref{f2w}) to $\tilde{v}$, we can write, for $t \geq 0$,
\begin{equation}
\label{f2w1}
\int_{Q_\varepsilon} (v\nabla w \, (-\Delta)^{1/2}w)(t,x) dx   \,
 \leq \, C\varepsilon^{1/2} |\tilde{v}(t)|_1 |w(t)|_{3/2}^2
 + \left|\int_{Q_\varepsilon} (v_3 \partial_{x_3}w
 (-\Delta)^{1/2}w)(t,x) dx \right|~.
\end{equation}
But an anisotropic H\"{o}lder inequality and Lemma \ref{lema2} imply
that, for $t \geq 0$,
\begin{equation*}
\begin{split}
\int_{Q_\varepsilon} (v_3 \partial_{x_3}w
(-\Delta)^{\frac{1}{2}}w)(t,x) dx \, & \leq \,
C  \varepsilon^{-\frac{1}{q}} \|v_3(t)\|_{L^q} 
\|(-\Delta)^{\frac{1}{2}}w(t)\|_{L^{\frac{2q}{q-1},2}}
\|\partial_{x_3}w(t)\|_{L^{\frac{2q}{q-1},2}}  \\
&\leq \, C \varepsilon^{-\frac{1}{q}} \|v_3(t)\|_{L^q} 
|w(t)|^2_{1 + 1/q}~,
\end{split}
\end{equation*}
or also, due to the Poincar\'{e} inequality for $w$,
\begin{equation}
\label{f2wv3}
\int_{Q_\varepsilon} (v_3 \partial_{x_3}w \,
(-\Delta)^{1/2}w)(t,x) dx  \,  \leq \,
C \varepsilon^{1-3/q} \|v_3(t)\|_{L^q} |w(t)|^2_{3/2}~.
\end{equation}
To estimate the third term, we again write $v$ as $\tilde{v} + v_3$,
apply the inequality (\ref{f3w}) to $\tilde{v}$ and remark that
$\int_{Q_\varepsilon} w\nabla v_3 ((-\Delta)^{1/2}w_3) dx =
-\int_{Q_\varepsilon} v_3 w (\nabla((-\Delta)^{1/2}w_3)) dx$, which
implies that, for $t \geq 0$,
\begin{equation}
\begin{split}
\label{f3w1}
\int_{Q_\varepsilon} (w\nabla v (-\Delta)^{1/2}w)(t,x) dx
& \,  \leq \, C\varepsilon^{1/2} |\tilde{v}(t)|_1 |w(t)|_{3/2}^2 +
\left|\int_{Q_\varepsilon} (w\nabla v_3 (-\Delta)^{1/2}w_3)(t,x)
dx\right| \\
&\, \leq \, C \Big( \varepsilon^{1/2} |\tilde{v}(t)|_1 |w(t)|_{3/2}^2 +
 |w_3(t)|_{3/2} \|(-\Delta)^{1/4}(wv_3)(t)\|_{L^2} \Big)~.
\end{split}
\end{equation}
It remains to bound the term $\|(-\Delta)^{1/4}(wv_3)\|_{L^2}$. A
quick computation using Fourier series shows that we have, for any 
$h \in \dot{H}^{1/2}_p(Q_\varepsilon)$, 
\begin{equation*}
\|(-\Delta)^{1/4}h\|_{L^2} \, \leq \,
\|\, \|(-\Delta_2)^{1/4}h\|_{L^2_{x'}(\Omega)}
\|_{L^2_{x_3}(0,\varepsilon)}
+ \|\, \|(-\partial^2_{x_3x_3})^{1/4}h\|_{L^2_{x_3}(0,\varepsilon)}
\|_{L^2_{x'}(\Omega)}~.
\end{equation*}
 Since $v_3$ is independent of $x_3$, it follows from the above 
 inequality that
\begin{multline}
\label{f3w2}
\|(-\Delta)^{1/4}(wv_3)\|_{L^2} \, \leq \, C \Big(
\|\, \|(-\Delta_2)^{1/4}(v_3w)\|_{L^2_{x'}(\Omega)} \|_{L^2_{x_3}
(0,\varepsilon)} \\
+ \|v_3 \|(-\partial^2_{x_3x_3})^{1/4}w\|_{L^2_{x_3}(0,\varepsilon)}
\|_{L^2_{x'}(\Omega)} \Big)~.
\end{multline}
But, the H\"{o}lder inequality, Lemma \ref{lema2} and the
Poincar\'{e} inequality (\ref{Poincarewbis}) imply that
\begin{equation}
\label{f3w3}
\|v_3 \|(-\partial^2_{x_3x_3})^{\frac{1}{4}}
w\|_{L^2_{x_3}(0,\varepsilon)}\|_{L^2_{x'}(\Omega)} 
\, \leq  \, C \|v_3\|_{L^q(\Omega)} 
\|(-\partial^2_{x_3x_3})^{\frac{1}{4}}w\|_{L^{2q/(q-2), 2}}
\, \leq \, C \varepsilon^{1- \frac{3}{q}} \|v_3\|_{L^q} 
|w|_{\frac{3}{2}}~.
\end{equation}
On the other hand, applying the periodic version of the multiplication
property given in Theorem 5.1 of \cite{Johnsen}, we can write,
\begin{equation*}
\begin{split}
\|\, \|(-\Delta_2)^{1/4}(v_3w)\|_{L^2_{x'}(\Omega)} \|^2_{L^2_{x_3}
(0,\varepsilon)} \, \leq \,
C (\int_0^\varepsilon ( & \|(-\Delta_2)^{1/4}v_3\|^2_{L^r(\Omega)}
\|w\|^2_{L^{2r/(r-2)}_{x'}(\Omega)} \\
&+ \|(-\Delta_2)^{1/4}w\|^2_{L^{2q/(q-2)}_{x'}(\Omega)}
\|v_3\|^2_{L^q(\Omega)}) dx_3 )~,
\end{split}
\end{equation*}
where $r=\frac{4q}{q+2}$. Using the two-dimensional Sobolev
embedding theorems, we infer from the above estimate that
\begin{equation}
\begin{split}
\label{f3w4}
\|\, \|(-\Delta_2)^{1/4}(v_3w)\|_{L^2_{x'}(\Omega)} \|^2_{L^2_{x_3}
(0,\varepsilon)} \, \leq \,
C (\int_0^\varepsilon ( & \|(-\Delta_2)^{1/4}v_3\|^2_{L^r(\Omega)}
\|(-\Delta_2)^{s_1/2}w\|^2_{L^2_{x'}(\Omega)} \\
&+ \|(-\Delta_2)^{1/4+ s_2/2}w\|^2_{L^2_{x'}(\Omega)}
\|v_3\|^2_{L^q(\Omega)}) dx_3 )~,
\end{split}
\end{equation}
where $s_1= 2/r$ and $s_2=2/q$. Applying then the following
Gagliardo-Nirenberg inequality (see
\cite{BahouriGallagher} or \cite{BerghLofstrom})
\begin{equation*}
\|(-\Delta_2)^{1/4}v_3\|_{L^r(\Omega)}
\, \leq \, C_q \|v_3\|^{1/2}_{L^q(\Omega)}
\|(-\Delta_2)^{1/2}v_3\|^{1/2}_{L^2(\Omega)}~,
\end{equation*}
we deduce from (\ref{f3w4}) that
\begin{equation}
\begin{split}
\label{f3w5}
\|\, \|(-\Delta_2)^{1/4}(v_3w)\|_{L^2_{x'}(\Omega)} \|_{L^2_{x_3}
(0,\varepsilon)} \, \leq \,
C_q ( & \|v_3\|^{1/2}_{L^q(\Omega)}
\|(-\Delta_2)^{1/2}v_3\|^{1/2}_{L^2(\Omega)}
\|(-\Delta)^{s_1/2}w\|_{L^2}\\
& + \|v_3\|_{L^q(\Omega)})
\|(-\Delta)^{\frac{s_2}{2}+ \frac{1}{4}}w\|_{L^2} ) \\
\, \leq \, C_q \Big( & 
\varepsilon^{\frac{3}{4}-\frac{3}{2q}}\|v_3\|_{L^q}^{\frac{1}{2}}
|v_3|_1^{\frac{1}{2}}
+ \varepsilon^{1-\frac{3}{q}}\|v_3\|_{L^q}\Big) |w|_{3/2}~.
\end{split}
\end{equation}
Finally, due to the estimates (\ref{f1w}), (\ref{f2w1}),
(\ref{f2wv3}), (\ref{f3w1}), (\ref{f3w2}), (\ref{f3w3}) and
(\ref{f3w5}), the solution $u_m=w_m +v_m$ satisfies the following
inequality, for $t \geq 0$,
\begin{multline*}
  \partial_t |w_m(t)|_{1/2}^2 + 2 \Big( \frac{3\nu}{4} -
  C_1|w_m(t)|_{1/2} - C_2 \varepsilon^{1/2} |\tilde{v}_m(t)|_1  -
 c_{1q}
\varepsilon^{1-3/q}\|v_{m3}(t)\|_{L^q} \\
  - c_{2q}
\varepsilon^{3/4-3/(2q)}\|v_{m3}(t)\|_{L^q}^{1/2}|v_{m3}(t)|_1^{1/2}
\Big) |w_m(t)|_{3/2}^2 \\
  \leq \, C_3\nu^{-1} \varepsilon \nl{(I-M)Pf}{2}^2~.
\end{multline*}
Since $u_m(t)$ belongs to the space $C^0([0,\tau);V^2_p \cap
\mathcal{V}_m)$, we infer from the properties (\ref{4Pm}),
(\ref{5Pm}), (\ref{5Pm3q1}) and the hypothesis (\ref{H1thq}) on the initial
conditions, where $k_1(q)$, $k_2(q)$, $k_5(q)$ are small enough,
that there exists a positive time $T$ such that, for $t \in [0, T)$,
\begin{equation}
\label{Hwvv3}
C_1|w_m(t)|_{\frac{1}{2}} + C_2 \varepsilon^{\frac{1}{2}} 
|\tilde{v}_m(t)|_1
+ c_{1q}\varepsilon^{1-\frac{3}{q}}\|v_{m3}(t)\|_{L^q} 
+c_{2q} \varepsilon^{\frac{3}{4}-\frac{3}{2q}}
\|v_{m3}(t)\|_{L^q}^{\frac{1}{2}}|v_{m3}(t)|_1^{\frac{1}{2}}
\, < \, \frac{\nu}{2}~,
\end{equation}
and, if $T < +\infty$,
\begin{equation}
\label{Hwvv3egal}
C_1|w_m(T)|_{\frac{1}{2}} + C_2 \varepsilon^{\frac{1}{2}} 
|\tilde{v}_m(T)|_1
+ c_{1q}\varepsilon^{1-\frac{3}{q}}\|v_{m3}(T)\|_{L^q} 
+c_{2q} \varepsilon^{\frac{3}{4}-\frac{3}{2q}}
\|v_{m3}(T)\|_{L^q}^{\frac{1}{2}}|v_{m3}(T)|_1^{\frac{1}{2}}
\, = \, \frac{\nu}{2}~,
\end{equation}
Then, as shown in the proof of Theorem \ref{theoreme2}, $w_m(t)$,
$\tilde{v}_m(t)$ and $v_{m3}(t)$ satisfy the estimates
(\ref{est1w}), (\ref{est2w}), (\ref{winte0t}), (\ref{intwfinal}),
(\ref{f4v}) and (\ref{v33}), for $t \in
[0,T]$. Moreover, we deduce from the inequalities (\ref{Lqv3}),
(\ref{intwfinal}) and the property (\ref{5Pm3q2}),
that, for $t \in [0,T]$,
\begin{equation}
\label{v3qest}
\|v_{m3}(t)\|_{L^q} \, \leq \, C(q) ( \|v_{03}\|_{L^q} +
\sup_s \|\nabla (-\Delta_2)^{-1}(M(Pf(s))_3)\|_{L^q} + \varepsilon^{-1/2 + 3/q}
D_1^{1/2})~.
\end{equation}
Using the estimates
(\ref{est1w}), (\ref{f4v}), (\ref{v33}) and (\ref{v3qest}), one shows
that, if
the hypotheses (\ref{H1thq}) and (\ref{H2thq}) are satisfied
for sufficiently small constants $k_1(q)$, $k_2(q)$, $k_3(q)$, $k_4(q)$,
 $k_5(q)$ and $k_6(q)$, then, for $t\in [0,T]$,
\begin{equation*}
C_1|w_m(t)|_{\frac{1}{2}} + C_2 \varepsilon^{\frac{1}{2}} 
|\tilde{v}_m(t)|_1
+ c_{1q}\varepsilon^{1-\frac{3}{q}}\|v_{m3}(t)\|_{L^q} 
+c_{2q} \varepsilon^{\frac{3}{4}-\frac{3}{2q}}
\|v_{m3}(t)\|_{L^q}^{\frac{1}{2}}|v_{m3}(t)|_1^{\frac{1}{2}}
\, < \, \frac{\nu}{4}~,
\end{equation*}
which contradicts the equality (\ref{Hwvv3egal}). It follows that
$T=+\infty$. Thus, we have proved that, under the hypotheses
(\ref{H1thq}) and (\ref{H2thq}), for every integer $m$, $m \geq m_0$,
the solution $u_m \in C^1([0, +\infty); \mathcal{V}_m)$ of the
modified Navier-Stokes equations
(\ref{NSabs}) with initial data $u_m(0)= \mathcal{P}_m u_0$
satisfies
\begin{equation}
\label{5inegamq}
\sup_{t \geq 0}\Big(|w_m(t)|_{\frac{1}{2}} + 
\varepsilon^{\frac{1}{2}} |\tilde{v}_m(t)|_1
+ \varepsilon^{1-\frac{3}{q}}\|v_{m3}(t)\|_{L^q} 
+ \varepsilon^{\frac{3}{4}-\frac{3}{2q}}
\|v_{m3}(t)\|_{L^q}^{\frac{1}{2}}|v_{m3}(t)|_1^{\frac{1}{2}}
\Big)\, < C~,
\end{equation}
where $C$ is a positive constant independent of $\varepsilon$
and $m$.
We now finish the proof, by arguing as in the proof of Theorem
\ref{theoreme2}.
\end{proof}

\providecommand{\bysame}{\leavevmode\hbox to3em{\hrulefill}\thinspace}

\vskip\baselineskip

\

\begin{minipage}[t]{7cm}
{\bf D. Iftimie}\\
IRMAR\\
Universit\'e de Rennes 1\\
Campus de Beaulieu\\
35042 Rennes Cedex (France)\\
{\sf Email: iftimie@maths.univ-rennes1.fr}
\end{minipage}
\hfill
\begin{minipage}[t]{8cm}
{\bf G. Raugel}\\
UMR 8628 \\
Math\'ematiques, B\^at. 425\\
Universit\'e de Paris-Sud\\
91405 Orsay Cedex (France)\\
{\sf Email: Genevieve.Raugel@math.u-psud.fr}

\end{minipage}

\


\end{document}